\documentclass[preprint]{aastex701}

\usepackage{longtable}
\usepackage{tabularx}
\usepackage{ltablex}   
\usepackage{booktabs}
\keepXColumns

\usepackage{amsmath}

\usepackage{xcolor}
\usepackage[normalem]{ulem}
\usepackage{hyperref}
\begin{document}

\title{The Role of Preflare Reconnection, Ideal MHD Instabilities, and Magnetic Nonpotentiality in Eruptive and Confined Solar Flares}

\author[0000-0001-8749-1022]{Wen He}
\affiliation{Institute for Space Weather Sciences, New Jersey Institute of Technology, University Heights, Newark, NJ 07102-1982, USA}
\affiliation{Center for Solar-Terrestrial Research, New Jersey Institute of Technology, University Heights, Newark, NJ 07102, USA}
\email[show]{wen.he@njit.edu}  

\author[0000-0002-8179-3625]{Ju Jing}
\affiliation{Institute for Space Weather Sciences, New Jersey Institute of Technology, University Heights, Newark, NJ 07102-1982, USA}
\affiliation{Center for Solar-Terrestrial Research, New Jersey Institute of Technology, University Heights, Newark, NJ 07102, USA}
\email{ju.jing@njit.edu}

\author[orcid=0000-0002-5865-7924, gname=Jeongwoo,sname=Lee]{Jeongwoo Lee}
\affiliation{Institute for Space Weather Sciences, New Jersey Institute of Technology, University Heights, Newark, NJ 07102-1982, USA}
\affiliation{Center for Solar-Terrestrial Research, New Jersey Institute of Technology, University Heights, Newark, NJ 07102, USA}
\email{leej@njit.edu}

\author[orcid=0000-0002-5233-565X, gname=Haimin,sname=Wang]{Haimin Wang}
\affiliation{Institute for Space Weather Sciences, New Jersey Institute of Technology, University Heights, Newark, NJ 07102-1982, USA}
\affiliation{Center for Solar-Terrestrial Research, New Jersey Institute of Technology, University Heights, Newark, NJ 07102, USA}
\email{haimin.wang@njit.edu}

\begin{abstract}
Identifying pre-flare magnetic conditions that determine whether a flare will be eruptive (associated with a coronal mass ejection) or confined remains a major challenge in solar physics. Motivated by previous statistical results showing that some eruptive flares occur under ideal MHD-stable conditions, we investigate the roles of kink instability (KI), torus instability (TI), preflare reconnection traced by precursor brightenings (PBs) and magnetic nonpotentiality in solar eruptiveness using an expanded sample of 68 flares above GOES class M4 from 2011--2024. To characterize the nonpotentiality of each active region (AR), we calculate the relative helicity $H_R$, current-carrying helicity $H_J$ and free magnetic energy. We find that, (1) 17 of 18 TI-unstable events and 12 of 19 KI-unstable events are eruptive, indicating that the TI parameter outperforms the KI parameter (2) although PBs are more commonly found in the MHD-unstable regime, only 3 of the 12 eruptive events in the TI/KI-stable regime show PBs, suggesting that preflare reconnection may play a secondary role compared with other factors. (3) The strongest indicator of eruptiveness is obtained by combining the TI parameter with magnetic nonpotentiality. Specifically, 24 of 36 eruptive flares satisfy $\sqrt{\left| H_J \right|}/\phi \geq 0.1$ or $n \geq 1.13$, whereas 24 of 25 confined flares satisfy $\sqrt{\left| H_J \right|}/\phi \le 0.1$ and $n \le 1.13$, where $\phi$ is the total unsigned magnetic flux and  $n$ is the decay index. These results suggest that the overall magnetic environment of ARs may play a significant role in determining the eruptiveness, in addition to the effect of MHD instability and reconnection. 
\end{abstract}


\section{Introduction}\label{0} 

Solar flares are energetic phenomena in the solar atmosphere and can be either eruptive or confined. Eruptive flares are  associated with coronal mass ejections (CMEs), while confined flares are not.  Understanding the physical conditions that determine whether a flare becomes eruptive or confined remains an important open question in solar physics.

One type of mechanisms proposed to explain the eruptiveness of solar flares involves ideal magnetohydrodynamic  (MHD) instabilities, including kink instability (KI) and torus instability (TI) of a magnetic flux rope (MFR). An MFR, characterized by twisted magnetic field lines wrapped around a common axis, is widely regarded as the core magnetic structure involved in solar eruptions, supported by observational, theoretical, and numerical studies. The KI and TI are associated with, respectively, excessive twisting of magnetic field lines about the MFR axis and the properties of the strapping  magnetic field above the MFR. Numerical studies have employed various analytical flux rope models as initial conditions in MHD simulations to investigate the evolution of the KI \citep{1999Titov, 2004Torok, 2003Fan}. When the twist of an MFR exceeds a critical threshold, which depends on the loop geometry and surrounding magnetic field, it becomes susceptible to the KI, leading to expansion and deformation. However, this evolution may quickly slow and remain locally confined. The MHD simulations by \cite{2005Torok} showed that a kink-unstable flux rope can evolve into a full eruption when the strapping magnetic field decreases sufficiently rapidly with height, suggesting the important role of the TI in determining whether a kink-unstable MFR can develop into a full eruption. The TI is commonly quantified by the decay index of the overlying strapping magnetic field. Originally introduced to describe the expansion instability of a toroidal current ring in laboratory experiments, the concept of the TI has since been applied to more general and complex cases such as solar eruptions, especially during their initial stage \citep{2006Kliem&Torok}. Notably, a laboratory experiment designed to study Sun-like line-tied MFRs under ideal MHD conditions identified four eruption regimes,  distinguished by whether the KI and TI parameters lie above or below their critical thresholds \citep{2015Myers}.

Although the roles of KI and TI in distinguishing confined and eruptive events have been demonstrated in MHD simulations and laboratory experiments, their effectiveness in actual solar eruptions remains less certain. For example, a statistical analysis of 38 major solar flares by \citet{2018Jing} failed to reproduce the four eruption regimes identified in laboratory experiments using the KI and TI parameters derived from solar observations. In that study, the KI parameter, magnetic twist number, shows little discriminating power between eruptive and confined events, and the TI parameter, decay index, exceeding a threshold is a sufficient but not necessary condition for eruption. Some MFRs located in both the TI- and KI-stable regimes still evolve into CMEs. Similar results were reported in another statistical study by \citet{2019Duan}. On the other hand, even in laboratory experiments, \citet{2015Myers} reported a special non-eruptive case, referred to as a ``failed torus'', in which an MFR located in the torus-unstable regime (characterized by a high TI parameter but a relatively small KI parameter) does not erupt  (see also \citealt{2019Zhou, 2021Alt}). Such ``failed torus'' events have not been identified  in observational statistical studies by \citet{2018Jing} and \citet{2019Duan}. These results suggests that KI and TI alone can not fully determine whether an MFR eruption succeeds or fails, and that additional mechanisms must be considered for solar eruptions. In addition to the local TI and KI-indices, the presence of filaments, the overall magnetic structure, as well as persistent pre-eruption heating and fast plasma flows may also influence the onset of eruptions \citep{Lee2016ApJ...831L..18L, Lee2018ApJ...856...70L}.

In particular, magnetic reconnection occurring prior to the main phase of flares provides an additional eruption mechanism and should be considered along with KI and TI. In particular, the tether-cutting reconnection below an MFR \citep{2001Moore, 2021Jiang} and breakout reconnection above the core magnetic field \citep{1999Antiochos, 2008Lynch} are two key reconnection processes that can facilitate the eruption during its early phase. Unlike the ideal MHD instabilities mentioned above, magnetic reconnection can occur in a resistive MHD process, where the frozen-in flux assumption is violated and magnetic topology is no longer preserved. For example, MHD simulations with flux emergence by \citet{2003Fan, 2007Fan} emphasized the role of reconnection during the nonlinear phase in forming twisted MFRs and triggering KI-driven eruptions.

One important observational signature of such pre-flare reconnection \textbf{(PFR hereafter)} is the occurrence of flare precursor brightenings (PBs). PBs are frequently observed as localized small-scale brightening preceding the flare main phase \citep{2015Wang, 2017Yumi, 2025Dissauer}. Although their precise role in large-scale flare energy release is still open to discussion, PBs are nontrivial considering its connection to the subsequent eruption process. For instance, magnetic reconnection signified by observed PBs may disturb the surrounding magnetic field and promote  breakout reconnection \citep{2004Sterling}, or generate sheared arcades that later enable tether-cutting reconnection \citep{2001Moore}. PBs may also be associated with magnetic flux emergence or flux cancellation that precedes the main reconnection episode \citep{2017Bamba, 2012Kusano}. Therefore, PBs provide valuable evidence of pre-flare magnetic reconnection and serve as an important diagnostic for understanding flare eruptiveness. More recently, high-resolution multi-wavelength observations have enabled detailed analysis of fine-scale magnetic structures in solar eruption events, further suggesting the contribution of small-scale PFR processes to solar eruptions. The observational evidence of fine-scale magnetic reconnection in the low corona with both imaging and spectropolarimetric data from advanced ground-based solar telescopes, such as the Goode Solar Telescope and the Daniel K. Inouye Solar Telescope, have also been reported in recent studies (e.g., \citealt{2025daSilva, 2025Mancuso}). 

The role of PFR in flare eruptions has also been explored in data-constrained MHD simulations with realistic observations. For example, \citet{2025Matsumoto} show that reconnection occurring prior to the flare can be crucial for the dynamics of MFRs in the eruption and enhance the TI . Motivated by the recent high-resolution observations of small scale magnetic flux emergence before the eruption, \citet{2025Inoue} perform a series of data-constrained MHD simulations with different pre-flare settings. Their results suggest that small-scale pre-flare magnetic flux can accelerate eruptions during the main phase by enhancing the growth of TI. 

In addition to TI, KI and PFR, the overall nonpotentiality of a solar active region (AR) also plays an important role in its flare and CME productivity. The global nonpotentiality of an AR is often quantified by parameters such as free magnetic energy, relative magnetic helicity and current-carrying helicity. While these parameters characterize different aspects of the complex three-dimensional (3D)  magnetic field, they are all proxies for the energy available to power eruptions. In particular, free magnetic energy $E_f$ is defined as the energy difference between the total magnetic field and its corresponding potential field. Relative magnetic helicity is another important physical quantity that describes the complex twist, shear and linkage of magnetic field lines. Based on a range of pre-flare parameters related to magnetic energy and helicity, \citet{2023Duan} show that the flux-normalized magnetic helicity and the ratio of free to potential magnetic energy in host ARs possess the strongest predictive capability for distinguishing between confined and eruptive flares. The evolution of AR nonpotentiality, including $E_f$, the magnetic helicity and their ratios, has also been investigated in other statistical studies \citep{2023Wangquan, 2025Thalmann}. These studies suggest that combining local instability parameters with measures of AR nonpotentiality can efficiently improve the prediction of flare eruptiveness. 

In this paper, we perform a statistical analysis of flare events to access the roles of various magnetic parameters in solar eruptions. Compared to previous statistical studies, we incorporate PFR into the analysis alongside TI, KI and global nonpotentiality parameters. For each event, we search for flare PBs as signatures of PFR. The KI and TI parameters, namely the magnetic twist number and decay index of an MFR, are derived from coronal magnetic field extrapolation based on observed photospheric vector magnetograms. The nonpotentiality parameters of ARs, including free magnetic energy, relative magnetic helicity and current helicity, are also derived. We then statistically examine the discriminating power of these parameters in distinguishing eruptive and confined events, as well as their relationships to CME acceleration. Section \ref{1} describes the data and different methods used in this study, Section \ref{2} presents the results, and Section \ref{3} provides a discussion and summary.

\section{Flare Sample Collection and Analysis Pipeline}\label{1} 

The flare sample is selected using criteria similar to those in \cite{2018Jing}. To increase the sample size, we extend the time range to 2024 and relax the flare threshold from GOES class M5 to M4. 
We note that historical GOES flare classifications are not fully homogeneous across the transition to the GOES-R series under the recent recalibration \citep{2026Machol}. In this study, we consistently adopt the re-calibrated science-quality L2 GOES XRS measurement for flare events after 2017 September 6. The corresponding flare magnitude is also re-calibrated based on the flare peak flux for 1-min average GOES-16/XRS-B data. As a result, this study includes 68 major flare events (GOES class M4 and above) that occurred within 45$^\circ$ latitude or longitude of disk center between February 2011 and March 2024. These events originate from 45 different solar ARs. 

Flares are classified as eruptive and confined flares based on their CME association, determined mainly from coronagraphic observations by Solar Heliospheric Observatory (SOHO)/LASCO and Solar TErrestrial RElations Observatory (STEREO), as well as remote-sensing observations of the low solar atmosphere from the Atmospheric Imaging Assembly \cite[AIA,][]{2012Lemen} onboard the \textit{Solar Dynamic Observatory (SDO)}. In total, the sample consists of 26 confined flares, and 42 eruptive flares.

\begin{figure}[h]
\centering
\includegraphics[width=0.75\textwidth]{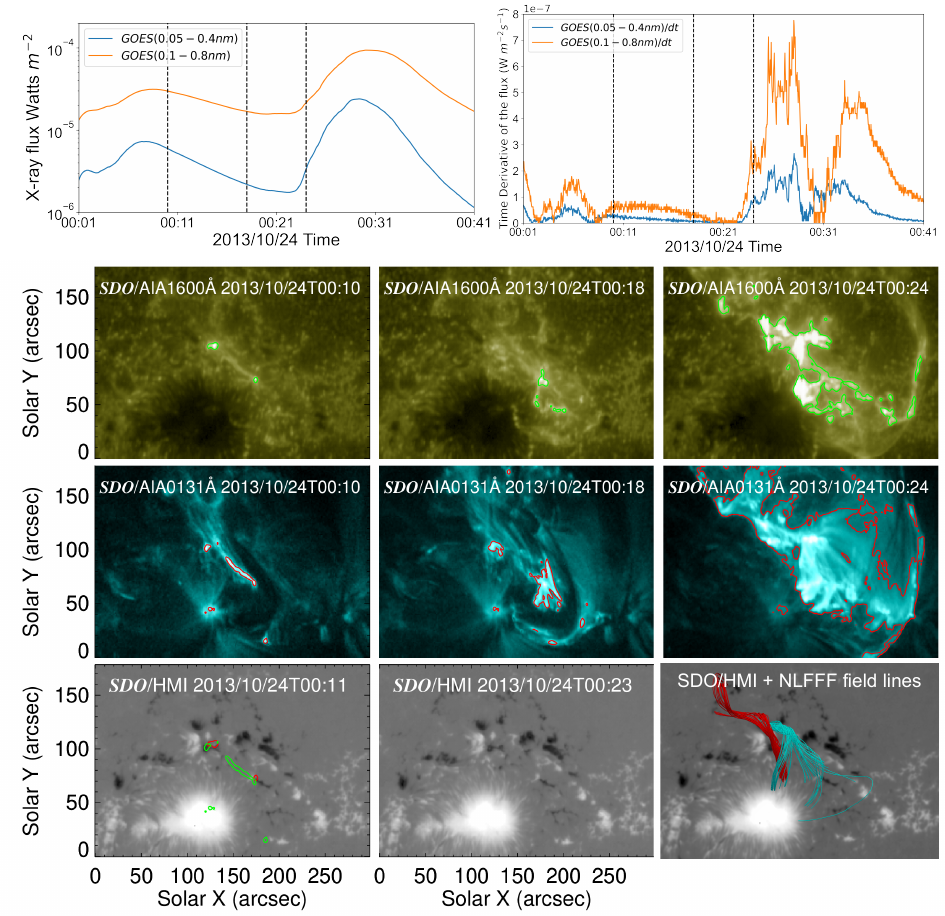}
\caption{Identification of the flare precursor brightening in the 2013/10/24T00:21 flare with GOES class M9.3. The top panels show the GOES lightcurves and their time-derivatives. The second and third rows show the evolution of brightening pixels in AIA 1600 \r{A} and 131 \r{A} channels, respectively. The fourth row shows the evolution of $B_{z}$ map from HMI SHARP vector magnetograms. The contours extracted from the intensity threshold (5 times of the standard deviation above the background intensity of AIA 1600 \r{A} and 131 \r{A} observations at 00:10) in red (1600 \r{A}) and green (131 \r{A}) colors are overplotted to corresponding AIA images, which indicate locations of the identified flare precursor. The contours from AIA 1600 \r{A} and 131 \r{A} channels obtained at 00:10 are also drawn over the lower left $B_z$ map. And the right figure in the fourth row illustrates two identified MFR structures before the flare onset based on the NLFFF result with overlaid HMI $B_z$ map at 00:11. The three dashed lines in the top panel indicate the selected time of AIA observations in the bottom panel.} \label{flare_precursor}
\end{figure}

\subsection{Identification of Flare PBs}

The identification of flare PBs requires the combined analysis of multi-wavelength observations and magnetic field data, as demonstrated in previous studies \citep{2007Chifor, 2014Awasthi, 2017Bamba}. Recent observations of PBs indicate that the associated soft X-ray emission often originates from lower atmospheric regions, such as footpoints and low-lying loop regions, distinguishing them  from emission during the main flare phase \citep{2021Hudson, 2025Dissauer}. 

To determine whether a flare exhibits PBs, we first examine the temporal association between the flare and preceding brightenings by inspecting soft X-ray light curves in 1-8 \r{A} and 0.5-4 \r{A} channels from GOES-15 and GOES-16. Specifically, we examine a time window spanning from two hours to four minutes prior to the flare onset to identify whether a gradual increase occurs before the main flare phase. When a noticeable enhancement is detected in the X-ray light curve, the time interval including the signal is selected for further analysis. Subsequently, EUV images from \textit{SDO}/AIA and co-aligned vector magnetograms from Helioseismic and Magnetic Imager \cite[HMI,][]{2014Bobra} onboard the \textit{SDO} are examined to determine the spatial association of the PBs with the flare location. For the consistency across the statistical study, AIA observations in 131 \r{A} and 1600 \r{A} channels are examined, and the magnetic polarity inversion line (PIL) derived from HMI amagnetogram is superimposed on the corresponding AIA images. An event is considered to exhibit a precursor if both temporal and spatial associations are satisfied, i.e., transient brightenings are observed within the pre-flare X-ray time interval, and are located near the major flaring PIL in both AIA channels. If another flare occurs within 30 minutes before the PB peak time or if the PB feature is weak near the major flaring PIL, we will classify it as an event with an uncertain PB. Figure \ref{flare_precursor} shows an example of PB identification for one flare event. The precursor enhancement in the soft X-ray light curve occurs about 15 minutes prior to the main flare, and the associated PBs are located along the magnetic PIL, coinciding with the site of the subsequent flare.

Based on our identification criteria, 36 out of 68 events exhibit clear PBs, 25 events show no detectable PBs, and 7 events are classified as uncertain due to ambiguous PB signatures. Note that the above identification method primarily captures PBs associated with tether-cutting reconnection occurring at the lower atmosphere, as these brightenings are localized near the flaring magnetic PILs. Breakout reconnection, however, often occurs at higher altitudes and may not be captured by this spatial criterion.

\subsection{NLFFF Extrapolation and MFR Identification} 

To derive magnetic parameters in three-dimensional space, we employ nonlinear force-free field (NLFFF) extrapolations based on photospheric vector magnetic field measurements. The force-free approximation is generally considered appropriate in the low-$\beta$ (the ratio between the gas pressure and magnetic pressure) solar atmosphere above the photosphere, where the magnetic pressure dominates over the gas pressure \citep{2001Gary}. We use pre-flare photospheric vector magnetograms from the HMI SHARP data product (\texttt{hmi.sharp\_cea\_720s}; \citealt{2014Bobra}) as the bottom boundary condition for the NLFFF extrapolations. The vector magnetograms are rebinned to a pixel scale of 1'' and preprocessed to reduce the net force and torque on the boundary, thereby better satisfying the force-free condition required for NLFFF extrapolation \citep{2006Wiegelmann}. The preprocessed data are then used as the bottom boundary for NLFFF extrapolation using the weighted optimization method \citep{2004Wiegelmann}. In this approach, the magnetic field is iteratively adjusted by minimizing a weighted optimization function that quantifies the force-free and divergence-free conditions until a converged solution is obtained. Over the past two decades, the weighted optimization method has been extensively evaluated using both analytical test cases and realistic photospheric magnetic field measurements, demonstrating high reconstruction accuracy, with reconstructed coronal fields differing from the reference fields by only a few percent under favorable conditions \citep{2009DeRosa, 2018Jing, 2021Wiegelmann, 2025Thalmann}.

\begin{figure}[h]
\centering
\includegraphics[width=0.9\textwidth]{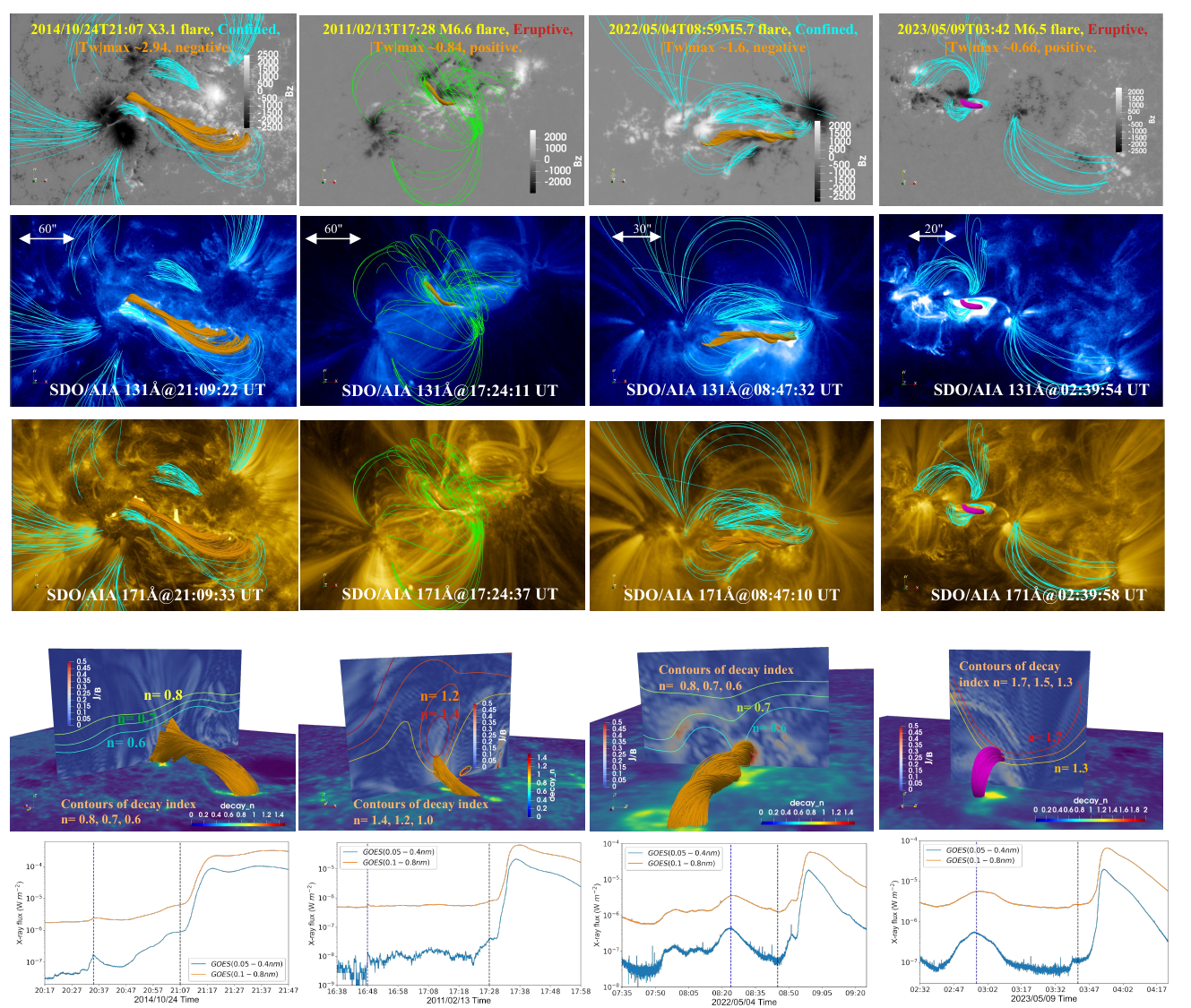}
\caption{Configurations of MFRs/sheared arcades and surrounding pre-flare magnetic field lines from NLFFF results. The first and second rows show MFR structures as well as selected field lines over the background of $B_z$ maps and AIA images, respectively. The second row shows the same MFRs overlaid with the coaligned $B_{z}$ map used for NLFFF extrapolations. The third row shows a distribution of $|\mathbf{J}| / |\mathbf{B}|$ on a vertical slice which crosses the identified MFRs or sheared arcades. The slice is also overlaid with contours of the decay index $n$. The bottom row shows GOES/X-ray lightcurves during the flare. The black dashed lines mark the flare onset time and the blue dashed lines indicate precursor enhancement before the flare which may be related to PB features prior to the main flare. } \label{mfr_aia_decay}
\end{figure}

A pre-existing MFR, or a strongly sheared sigmoidal structure, is generally required in eruption models in which the loss of ideal MHD stability triggers the eruption. In this study, the TI and KI parameters are evaluated only for events in which such structures can be clearly identified. We follow a procedure similar to that described by \citet{2022He} to identify MFRs or highly sheared arcades. First, we examine the evolution of AIA observations in the flaring region to find key observational features for each flare event. We then track groups of coherent magnetic field lines from the NLFFF extrapolation that exhibit relatively large magnetic twist values, without imposing a fixed threshold such as  $|{T_{w}}| = 1$, and compare them with the major flare signatures observed in AIA images, including hot channels and flare ribbons. We also calculate the length of magnetic field lines for each NLFFF extrapolation. Anomalously long field lines whose morphology and spatial extent are inconsistent with the corresponding coronal structures in AIA observations are excluded, while the dominant population of twisted field lines with a coherent length distribution and good morphological correspondence with observed coronal configuration is retained. 
Finally, a coherent, rope-like structure consistent with the AIA observations is chosen as the core magnetic structure for the subsequent analysis of the TI and KI parameters. Examples of MFRs identified from the NLFFF extrapolations and compared with the corresponding AIA observations are shown in Figure \ref{mfr_aia_decay}.

Given the non-uniqueness of NLFFF solutions among various NLFFF models \citep{2009DeRosa}, it is important to assess the quality and reliability of the extrapolated NLFFF models. We evaluate the force-free metric $CWsin$ and the divergence-free metric $|\langle f_{i} \rangle|$ of each NLFFF model, to ensure the solution is converged and consistent with the force-free conditions. For all NLFFF extrapolations used in this study, $ 0 <\textit{CWsin} < 0.3$ and $|\langle f_{i} \rangle| < 1.6 \times 10^{-4}$, which are comparable to the quality metrics reported in many previous NLFFF modeling results based on realistic photospheric vector magnetograms \citep{2009DeRosa, 2025Thalmann}.

In addition to these quantitative quality metrics, we compare extrapolated field lines with coronal structures observed by AIA as a complementary qualitative consistency check. For each flare event, we select several representative AIA images at different evolutionary stages, including the pre-flare phase, near the flare onset, and during the flare, to better identify and trace the evolution of major coronal structures. The selected AIA images are coaligned with the input HMI SHARP magnetograms, and the extrapolated field lines are  compared with observed features such as coronal loops, hot channels, localized brightenings. 

The second and third rows in Figure \ref{mfr_aia_decay} show representative comparisons between AIA observations and the corresponding NLFFF lines. Here, AIA 131 \r{A} observations are primarily used to identify hot core structures and coronal loops more closely associated with strongly non-potential magnetic configuration, while AIA 171 \r{A} images provide an additional comparison with the surrounding, relatively cooler coronal loops. In general, the reconstructed field lines show good morphological agreement with the observed coronal structures, particularly with the MFR-related core structures. Some deviations are present for the surrounding loops, especially in AIA 171 \r{A}. An exact one-to-one correspondence between individual extrapolated field lines and EUV loops is not expected, because AIA records optically thin, line-of-sight-integrated emission from plasma within temperature ranges determined by the response of each AIA passband. Consequently, only a subset of the magnetic structures may be visible in a given AIA channel \citep{2005Wiegelmann, 2017Duan}. Projection effects, finite spatial resolution, and uncertainties inherent in the NLFFF reconstruction can further contribute to apparent discrepancies between the modeled field lines and observed coronal structures. We therefore use the comparison with AIA observations as a complementary qualitative assessment rather than assigning a quantitative uncertainty to each NLFFF solution based on the apparent field-line alignment. If an NLFFF solution shows poor correspondence with AIA observations, especially in the major flaring region, that solution is excluded from the analysis. 

In the initial quality assessment of NLFFF results, three events, the M5.3 flare on 2012 July 4, the M7.3 flare on 2014 April 18, and the X1.4 flare on 2012 July 12, exhibited poor agreement between NLFFF lines and the corresponding AIA observations. We therefore excluded them from further analysis.  After removing those incompatible NLFFF results, 68 events remained in the final sample are listed in Table \ref{flare_event_list}. In this flare collection, MFRs are successfully identified in 61 out of 68 events, based on the combined NLFFF and AIA comparison. The comparison results for all analyzed events are available in the Zenodo repository\dataset[doi:10.5281/zenodo.23145233]{https://doi.org/10.5281/zenodo.23145233}. The KI and TI parameters are derived and analyzed for these events, as described in Section \ref{sec:TI_KI_parameters}. Among these 61 events, 11 contain two MFR structures, as one example shown in the second column of Figure \ref{mfr_aia_decay}. In such cases, we choose the MFR with the higher altitude for statistical analysis, following \citet{2019Duan}.

The remaining 7 events lacking identifiable MFRs are marked with ``NA'' in columns $n_{max}$,  $\left| n\right|$, $T_w$ and $\left| T_w \right|$ for Table \ref{flare_event_list}. For the other seven flare events, although their corresponding NLFFF field lines generally align with coronal loops in AIA observations, there are no twisted field line groups (with $\left| T_w \right| > 0.5$) found near flaring sites. The lack of twisted magnetic structures may be attributed to two factors. First, a coherent MFR may not yet have formed at the time of the NLFFF extrapolation. Second, the ability of current NLFFF models to reconstruct the coronal magnetic field can be limited by using the photospheric vector magnetograms as the bottom boundary. The photosphere is generally not force-free which is often preprocessed to meet the force-free conditions. Specifically, in these seven cases, comparisons between magnetic topology and AIA observations indicate specialized topological features are found near the major flaring areas, such as null points, quasi-separatrix layers (QSLs), and hyperbolic flux tubes.

\subsection{KI and TI Parameters of MFR Configurations} \label{sec:TI_KI_parameters} 
The KI and TI parameters are commonly represented by the magnetic twist number $T_w$ and the decay index $n$, respectively, and are derived from NLFFF and potential field models. Specifically, the magnetic twist number $T_{w}$, is defined by \citep{2006Berger, 2016LiuR},
\begin{equation}\label{eq:tw}
T_{w} = \int_{L} \frac{\mu_{0} J_{||}}{4\pi B} \,dl = \frac{1}{4\pi}\int_{L} \frac{(\nabla \times \textbf{B})\cdot \textbf{B} }{B^{2}} \,dl
=\frac{1}{4\pi}\int_{L} \alpha \,dl,  
\end{equation}

\noindent where $\alpha$ is the force-free parameter, i.e. $\nabla \times \textbf{B} = \alpha \textbf{B}$  and the integral is taken along an individual magnetic field line of length $L$. $T_{w}$ provides a good approximation of the winding number for two infinitely close field lines. 

We also compute another parameter, the squashing degree to examine the magnetic connectivity around MFRs. The squashing degree, also called $Q$ factor, is calculated based on the mapping for two points on a same field line to measure the distortion of the elemental magnetic flux tube \citep{1996Demoulin,2002Titov}.
The $Q$ factor is also useful to find regions with strong variation of magnetic field lines, the quasi-sepatrix layers (QSLs). 
The $Q$ factor for the field line connecting two footpoints is defined by \citep{2002Titov},
\begin{equation} \label{eq:Q}
    Q = \frac{a^2+b^2+c^2+d^2}{det(D_{12})}=\frac{a^2+b^2+c^2+d^2}{\left| ad - bc \right|}
\end{equation}
where the Jacobian matrix, $D_{12}$, for the mapping is calculated using a pair of footpoints ($x_1$, $y_1$) and ($x_2$, $y_2$), 
$$
D_{12} = \left[ \frac{\partial \textbf{r}_2}{\partial \textbf{r}_1} \right]= \begin{pmatrix}
\partial x_2 /\partial x_1 & \partial x_2 /\partial y_1\\ 
\partial y_2 /\partial x_1 & \partial y_2 /\partial y_1\\ 
\end{pmatrix} = \begin{pmatrix}
    a& b \\
    c& d \\
\end{pmatrix}.
$$
In this study, the twist number $T_w$ and squashing factor $Q$ are calculated using the code developed by \citet{2016LiuR} based on the NLFFF models.

On the other hand, the TI parameter is quantified by the decay index $n$,
\begin{equation} \label{eq:decay}
n = \frac{\partial \log(B_{ext})}{\partial \log(h)}, 
\end{equation}
\noindent which measures the rate at which the horizontal component of the overlying magnetic field $B_{ext}$ decreases with height $h$. The decay index is calculated from the potential field rather than the NLFFF to characterize the external background field.

\citet{2006Kliem&Torok} show that TI can trigger an eruption when the decay index directly above an MFR exceeds a critical threshold (typically $n >1.5$). Owning to the complex geometry of an MFR and the associated variations in magnetic tension and hoop forces, the decay index varies along the MFR axis  \citep{2019Duan}. To define a representative value for each event, we construct a decay index map and select a vertical slice approximately perpendicular to the MFR axis through the middle point of the MFR axis, as shown in Figure \ref{mfr_aia_decay}.

For this statistical study, we calculate single values of $T_w$ and $n$ for each individual flux rope. In reality, these parameters vary within and around a given flux rope given its topology, so the derived values across statistical studies can vary due to different definitions and methods. For example, \citet{2018Jing} compute $T_w$ as the mean unsigned value $\left|T_{w}\right|$ over MFR field lines, whereas \citet{2019Duan} adopt the maximum unsigned value ${\left|T_{w}\right|}_{\max}$, which is often regarded as a proxy for the MFR axis, as the controlling parameter. For $n$, both \citet{2018Jing} and \citet{2019Duan} evaluate $n$ at the apex height of the MFR, but in different coordinate systems. In this study, we use the mean unsigned twist number $\left|T_{w}\right|$ from identified MFRs (when present) to characterize the KI parameter, and adopt the maximum value of $n$ along MFR-associated field lines to represent the TI parameter. 

\subsection{Additional Magnetic Parameters} \label{sec:additional_mag_parameters}

To complement the MHD instability parameters of major magnetic structures prior to flares, we compute additional quantities to quantify the nonpotentiality of source ARs. The total magnetic energy and free magnetic energy are calculated within a main 3D computational domain V which is constructed for NLFFF extrapolations, 
\begin{equation}\label{eq:efree}
    E_{free} = E_{tot} - E_{pot} =  \int_V \frac{B_{nlfff}^2}{8\pi} dV - \int_V \frac{B_{pot}^2}{8\pi} dV 
\end{equation}

Magnetic helicity provides another way to quantify the complexity of magnetic topology, including the magnetic twist and mutual linkage between different magnetic field lines. To overcome the gauge freedom in the vector potential involved, \cite{1984Berger&Field} and \citet{1985Finn&Antonsen} introduced the concept of relative magnetic helicity, defined under the conditions that \textbf{B} is solenoidal and $\textbf{B}\cdot \textbf{n} = \textbf{B}_{p}\cdot \textbf{n}$ on the boundary. Under these conditions, the gauge-invariant relative helicity in 3D finite domains can be written as,
\begin{equation}\label{eq:hv}
   H_R =  \int_{V} (\textbf{A} + \textbf{A}_{p}) \cdot (\textbf{B} - \textbf{B}_{p}) dV,  
\end{equation}
where $\textbf{A}$ and $\textbf{A}_{p}$ are the vector potentials. We determine them from the NLFFF magnetic field \textbf{B} and the reference magnetic field $\textbf{B}_{p}$ (the  a potential field in this case) by solving $\textbf{B} = \nabla \times \textbf{A}$ and $\textbf{B}_{p} = \nabla \times \textbf{A}_{p}$, respectively. To simplify the calculation of the magnetic vector potential, we use the DeVore's gauge in which $z$-component of the vector potential is set to zero, i.e., $\hat{\textbf{z}} \cdot \textbf{A} = \hat{\textbf{z}} \cdot \textbf{A}_{p} =0$ \citep{2000DeVore}.
In addition, another gauge-invariant helicity $H_J$ is also examined to estimate the contribution from current-carrying system, in the form of:
\begin{equation}\label{eq:hj}
   H_J =  \int_{V} (\textbf{A} - \textbf{A}_{p}) \cdot (\textbf{B} - \textbf{B}_{p}) dV,  
\end{equation}
These nonpotentiality parameters, $E_f$ and $H_{R,J}$, (equations [\ref{eq:efree}--\ref{eq:hj}]) includes integration of the magnetic quantities over the whole volume of a given AR, whereas the former MHD instability parameters, $T_w$ and $n$, (equations [\ref{eq:tw}--\ref{eq:decay}]) are local variables. As another difference, $E_f$ and $H_{R,J}$ have no known theoretical threshold for the eruption. Only an empirical threshold for the eruption has been explored in terms of the ratio of current-carrying helicity to total relative helicity in numerical simulation \citep[e.g.][]{2026Bian}. 

\section{Results} \label{2}

\subsection{Event Classification}
\begin{figure}[h]
\centering
\includegraphics[width=0.35\textwidth]{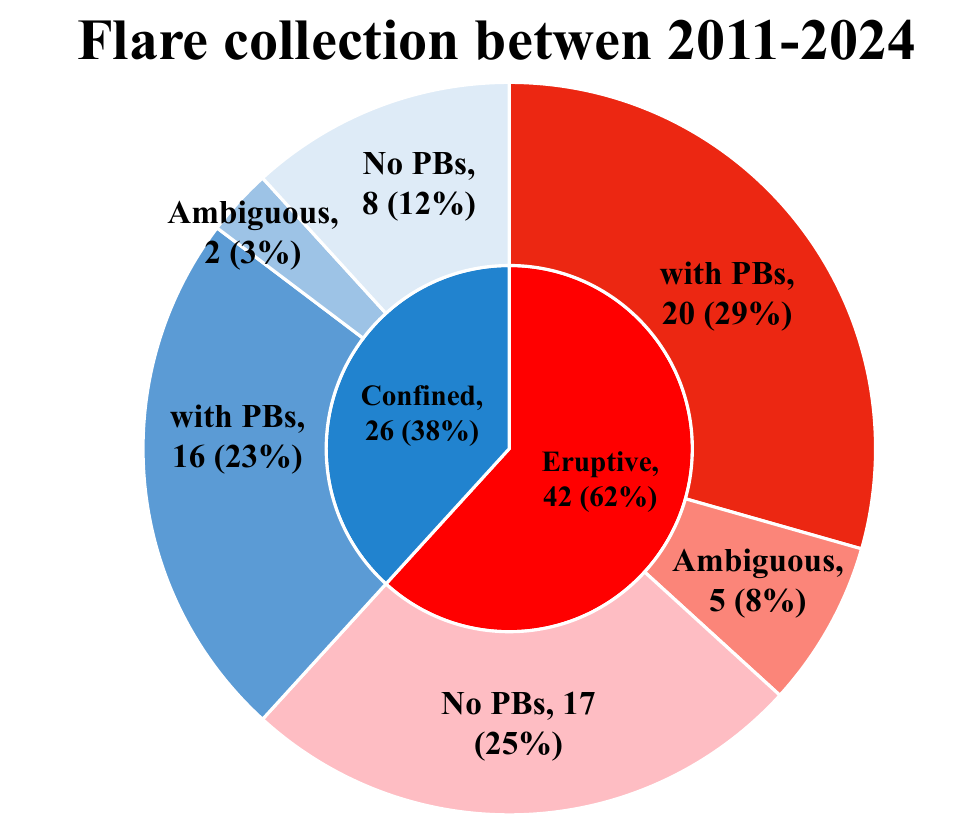}
\caption{Classification of the 68 flare events. The inner pie chart shows the division of events into eruptive and confined flares, while the outer pie chart further subdivides each category according to the PB identification.} \label{flare_sample}
\end{figure}

Figure \ref{flare_sample} presents an overview of the classification for the 68 selected flare events, and their key derived physical properties are summarized in Table \ref{flare_event_list} from Appendix. Among these events, 42 (62\%) flares are eruptive, 26 (38\%) are confined. From the perspective of PFR, 36 (53\%) events exhibit well-separated PB features, 25 (37\%) show no clear PB signatures, and 7 (10\%) are classified as ambiguous. As shown in Figure \ref{flare_sample}, PBs are more frequently found in confined events, while eruptive events show a more balanced distribution between cases with and without PBs. Table \ref{flare_event_list} lists the TI parameter, represented by the maximum decay index $n_{\rm max}$, the averaged decay index, $\left| n \right|$, 
derived from the identified MFRs, and the KI parameter, quantified by the average twist number $T_{w}$ of the MFRs as well as the maximum $T_w$ along the axis. In addition, it includes other quantities, such as the total AR magnetic flux, the ratio between magnetic free energy and magnetic potential energy, and the relative helicity and the current-carrying helicity, all computed over the entire extrapolation domain. 

\begin{figure}[h]
\centering
\includegraphics[width=0.6\textwidth]{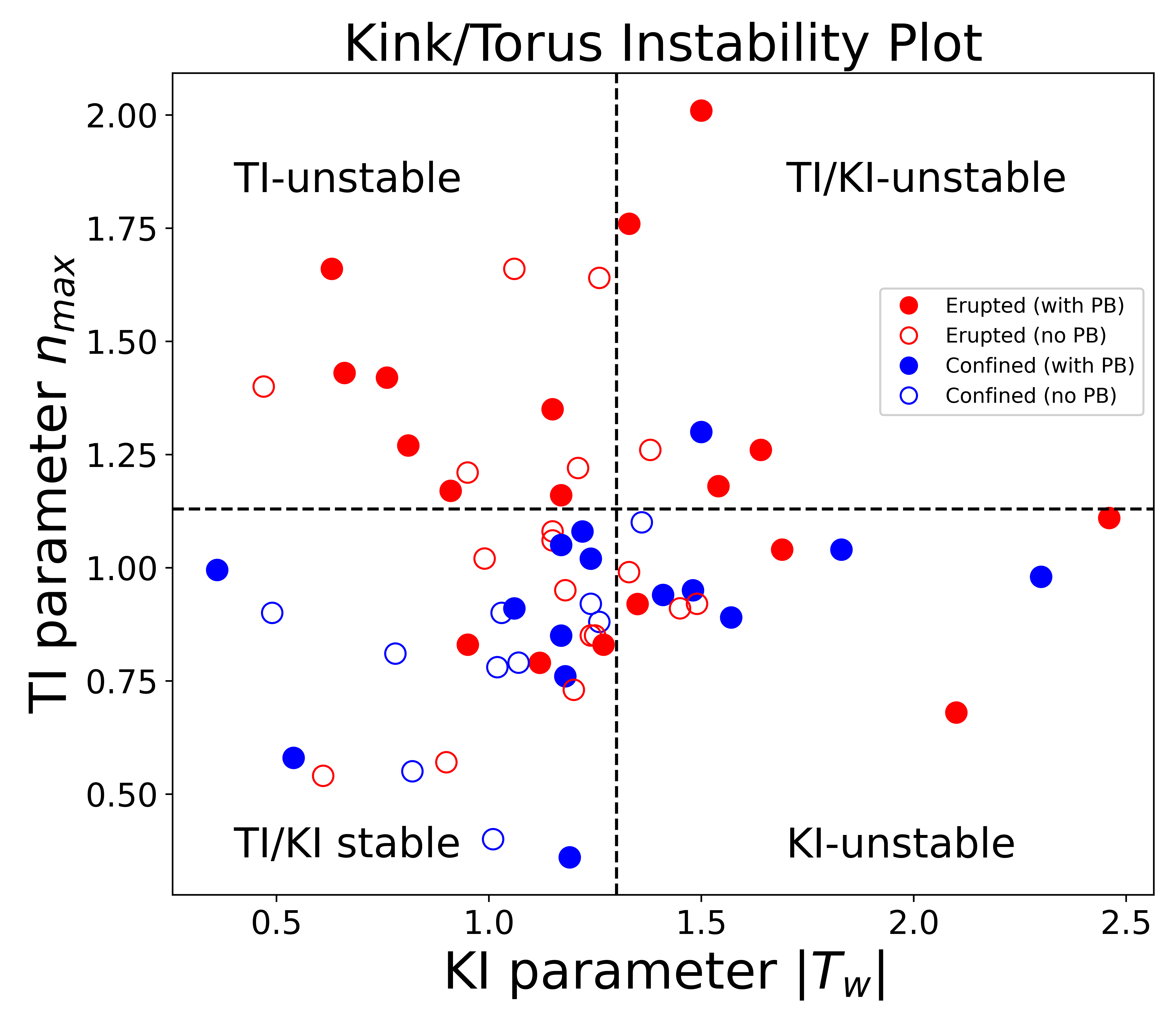}
\caption{The scatter plot between KI parameter $\left|T_{w} \right|$ and TI parameter, $n_{max}$ for the identified MFRs. 
The eruptive events are marked in red and confined events are in blue with the presence/absence of PB features denoted by filled/open circles. Four regimes are divided by the critical values $\left|T_{w}\right| = 1.3$ and $n_{\rm max} = 1.13$.} \label{KI_TI_map1}
\end{figure} 

\subsection{TI, KI and Reconnection} \label{subsec: ti_ki}
Figure \ref{KI_TI_map1} presents the scatter plot of the TI parameter $n_{\rm max}$ versus the KI parameter $\left|T_{w}\right|$ for the 61 events with identified MFRs. Eruptive and confined events are shown in red and blue, respectively, while PB signatures are indicated by filled (with precursors, suggesting PFR) and open (without precursors) circles. In the previous statistical studies for solar flare collections, different critical value for KI and TI parameters are adopted in \citet{2018Jing} ($n_{crit} \simeq 0.8$, no estimation for $\left|T_{w}\right|_{crit}$) and \citet{2019Duan} ($n_{crit} = 1.3$ and $\left|T_{w}\right|_{crit} = 2$). Based on the analytical analysis for stability of flux rope eruption, \citet{2010Demoulin} provide an approximation of $n_{crit}$ in the range of [1.1, 1.3]. So we empirically adopt critical thresholds $\left|T_{w}\right| = 1.3$ and $n_{\max} = 1.13$ to partition the TI-KI parameter space into four regimes: both TI/KI-unstable, both TI/KI-stable, TI-unstable, and KI-unstable regimes, analogous to the laboratory results \citep{2015Myers}. From this scatter plot, we note the following.

(1) When the KI parameter exceeds its threshold, i.e., within the KI-unstable and TI/KI-unstable regimes, about 63\% (12 out of 19) of events are eruptive. In contrast, when the TI parameter exceeds its  threshold, i.e., within the TI-unstable and TI/KI-unstable regimes, nearly 94\% of events (17 out of 18) are eruptive. This suggests that the TI parameter $n_{\rm max}$, when exceeding a critical threshold, is a strong indicator of eruptions, whereas the KI parameter $\left|T_{w}\right|$ plays a comparatively minor role.

(2) Within the TI/KI-stable regime, 40\% (12 out of 30) of events are nevertheless eruptive. Ideal MHD instabilities alone cannot account for these eruptions, and PFR, identified by PBs, is expected to act as an additional driver. However, only three of these 12 events exhibit PBs prior to eruption, and the remaining 9 events erupt without clear reconnection signatures. In addition, nine events remain confined despite evidence of reconnection. This suggests that PFR may facilitate eruption in some cases, yet it is neither necessary nor sufficient. 

(3) PFR, as manifested by PBs, occurs in both eruptive and confined flares as shown in Figure \ref{flare_sample}. Among the 37 events with PBs, 54\% are eruptive and 46\% are confined flares, further indicating that reconnection alone has limited discriminating power between eruptive and confined events. From a regime-based perspective, however, events with PBs occur more frequently in the TI-unstable, KI-unstable, and TI/KI-unstable regimes. Specifically, within the TI/KI-stable regime, only 40\% (12/30) of events exhibit PFR, whereas the fraction increases to about 69\% (9/13), 58\% (7/12) and 83\% (5/6) in the KI-unstable, TI-unstable and TI/KI-unstable regimes, respectively. 

(4) To further assess the role of PFR in eruptions, we examine the eruptive fraction among events with PBs in each regime. In the TI/KI-stable, KI-unstable, TI-unstable and TI/KI-unstable regimes, 25\% (3/12), 50\% (4/8), 100\% (7/7) and 80\% (4/5) of PB-associated events are eruptive, respectively. These results suggest that, for events with PFR, the likelihood of becoming eruptive events is low when the TI parameter is below its threshold, but increases substantially once the TI threshold is exceeded.

Taken together, these results indicate that PFR occurs more frequently under favorable instability conditions (i.e., within TI-unstable, KI-unstable and TI/KI-unstable regimes), but is strongly associated with eruptive outcomes only when TI condition is satisfied. This underscores the dominant role of TI in enabling eruptions. PFR therefore likely plays a secondary, facilitating role, in conjunction with TI.

\begin{figure}[h]
\centering
\includegraphics[width=0.4\textwidth]{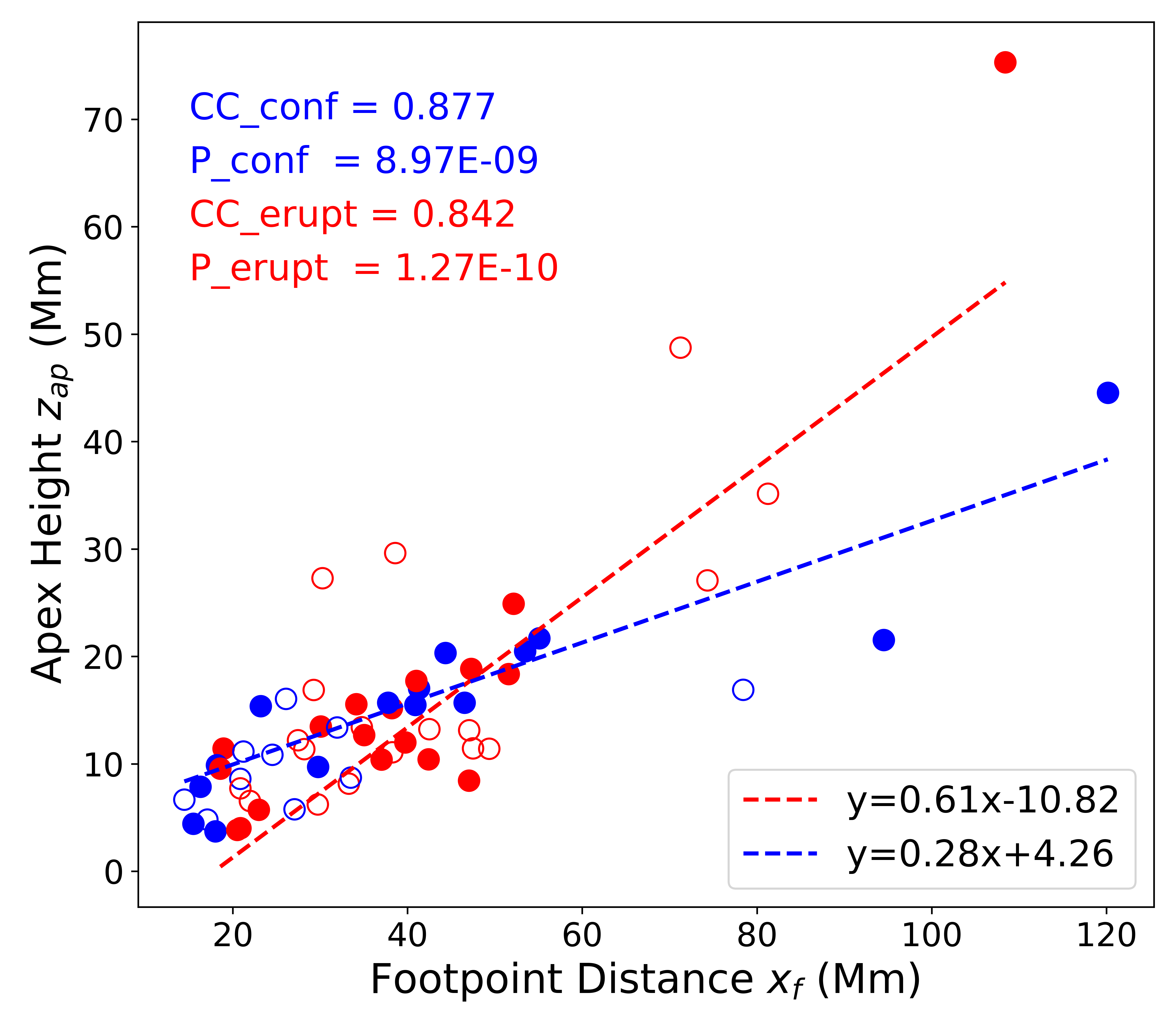}
\caption{The scatter diagram of the apex height $z_{ap}$ vs the footpoint distance $x_f$ of MFRs. The blue and red dashed lines indicate the linear fits for data points in eruptive events and confined events, respectively. For each event, the same convention for the symbols as in Figure \ref{KI_TI_map1} is used to represent the eruptiveness and PB occurrence.} \label{mfr_geometry}
\end{figure}

Based on laboratory experiment results, \citet{2021Alt} showed that the critical decay index $n_{cr}$ is not universal for each events as it also depends on the specific MFR geometry and physical properties, such as the normalized apex height (ratio between the MFR's apex height $z_{ap}$ and the footpoint distance $x_{f}$). The effect of the local critical decay index threshold has also been considered in a statistical study for the observations and NLFFF extrapolation results of eruptive filaments by \citet{2026ZhangC}. To assess the potential of  MFR geometry in determining the flare eruptiveness, we calculate the apex height $z_{ap}$ and footpoint distance $x_{f}$ for the axes of all identified MFRs.  Figure \ref{mfr_geometry} shows the scatter diagram of $z_{ap}$ -- $x_f$ for MFRs. Most MFRs are located at relatively low height (e.g., $z_{ap} < 25$ Mm), while some MFRs, more from eruptive events, reach heights above 25 Mm. Overall, the $z_{ap}$ versus $x_f$ diagram shows a clear positive correlation due to a common geometric scaling between height and distance. However, the pre-flare MFR geometry alone provides limited discriminating power in separating eruptive and confined events. Nevertheless, subtle systematic differences between the two groups can still be identified. The linear Pearson correlation coefficients (CCs) are 0.842 (p-value = $1.27\times 10^{-10}$) and 0.877 (p-value = $8.97\times 10^{-9}$) for eruptive and confined events, respectively, indicating greater geometric variability in eruptive events. In addition, eruptive events show a systematically higher height-to-distance ratio than confined events (0.61 versus 0.28), suggesting that eruptive MFRs are more elevated relative to their distance. 

\begin{figure}[h]
\centering
\includegraphics[width=0.8\textwidth]{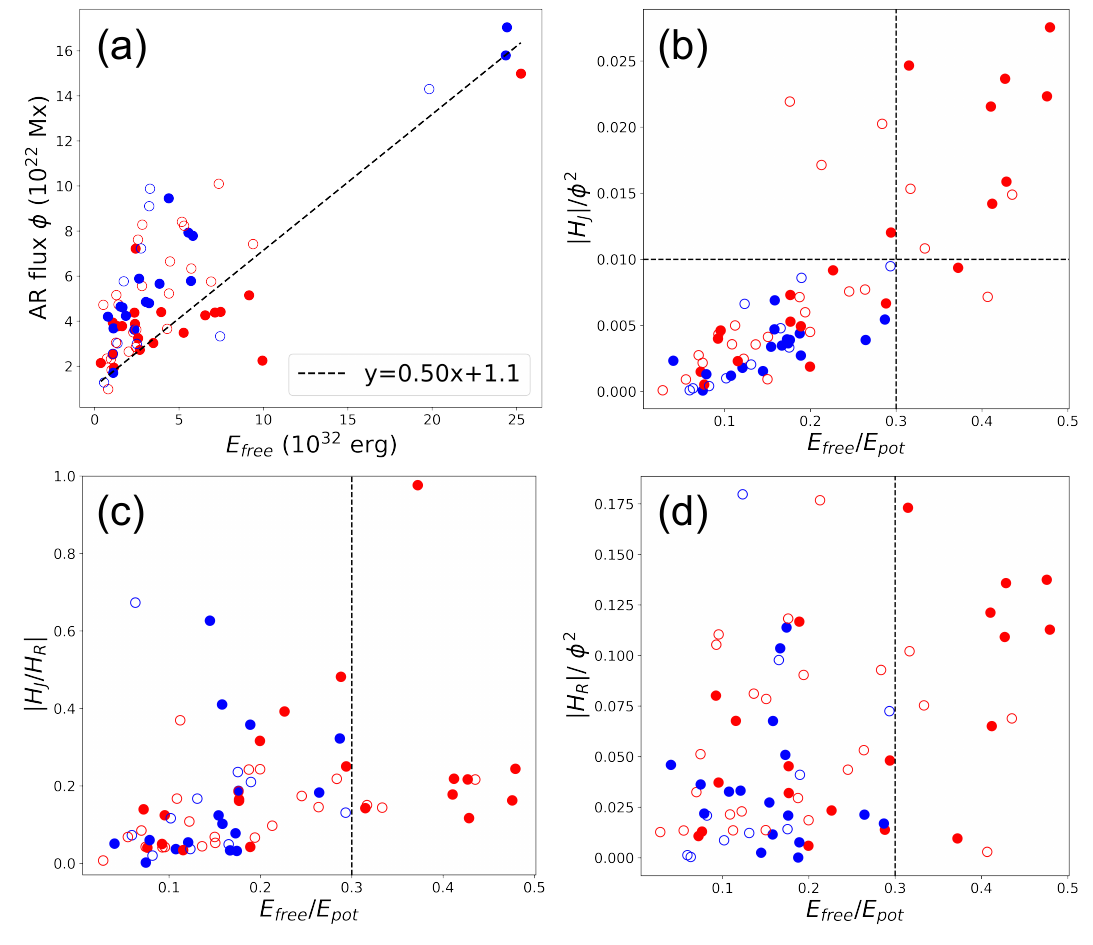}
\caption{Scatter plots of (a) AR flux vs $E_{free}$, (b) $\vert H_J\vert / \phi^2$ vs $E_{free} /E_{pot}$, (c) $\vert H_J / H_R \vert$ vs $E_{free} /E_{pot}$ and (d) $\vert H_R \vert/ \phi^2$ vs $E_{free} /E_{pot}$ based on Table \ref{flare_event_list}. The same symbol convention as in Figure~\ref{KI_TI_map1} is used to indicate the eruptiveness and PB occurrence of each event. In panel (a), the dashed line shows the linear fit to all data points. In panel (b), the whole regime is divided empirically by a horizontal dashed line ($\vert H_J \vert/ \phi^2 = 0.01$) and a vertical dashed line ($E_{free} /E_{pot} = 0.3$). The same vertical dashed line is also shown in (c) and (d). } \label{helicity_energy}
\end{figure}

\subsection{Magnetic Nonpotentiality} \label{subsec: nonpontentiality}

To study the role of the magnetic nonpotentiality in host ARs, we plot additional parameters in Figure \ref{helicity_energy}. The nature of these scatter plots are different from that of Figure \ref{KI_TI_map1}. While KI and TI parameters are localized quantities and are independent of each other, the parameters related to magnetic nonpotentiality may exhibit correlation with another to some extent. The scatter plots in Figure \ref{helicity_energy} can help reveal such relationships. Figure \ref{helicity_energy}a shows the scatter plot of AR flux versus free magnetic energy $E_{free}$. Both quantities are extensive parameters that increase with the AR size. Therefore, a positive correlation (Pearson CC = 0.80) between them is expected. The dashed line represents the best linear fit to all data points. We notice that most confined events lie above the fitted line, suggesting that confined events tend to occur in relatively larger but more potential ARs.

We also calculate the normalized free magnetic energy and helicity budgets within the modeling domain, to characterize the nonpotentiality and complexity of ARs prior to flares.  Figure \ref{helicity_energy}b -- \ref{helicity_energy}d shows scatter plots of $\vert H_J \vert/\phi^2$,  $\vert H_J/H_R \vert$, and $\vert H_R \vert/\phi^2$ versus $E_{free}/E_{pot}$. Among these, the $\vert H_J \vert/\phi^2$ versus $E_{free}/E_{pot}$ parameter pair (Figure \ref{helicity_energy}b) shows the strongest discriminatory power between eruptive (red) and confined (blue) events: events with $\vert H_J \vert/\phi^2 > 0.01$ and $E_{free}/E_{pot} > 0.3$ are exclusively eruptive, while confined events cluster in the regime $\vert H_J \vert/\phi^2 < 0.01$ and $E_{free}/E_{pot}< 0.3$. In contrast, the $\vert H_R \vert/\phi^2$ versus $E_{free}/E_{pot}$ and $\vert H_J/H_R \vert$ versus $E_{free}/E_{pot}$ diagrams (Figure \ref{helicity_energy}d and Figure \ref{helicity_energy}c) show less distinct separation, although most confined events remain concentrated in a narrower range with $E_{free}/E_{pot} < 0.3$. Despite differences in the flare sample and the extrapolation method, our findings are consistent with \citet{2023Duan}.

In addition, we find that in the highly nonpotential regime, events with PBs (filled circles) significantly outnumber those without PBs (open circles). For example, when $E_{free}/E_{pot}>0.3$ and $\vert H_J \vert/\phi^2 > 0.01$, the ratio of PB to non-PB events is 2, compared to 1.1 in the weakly nonpotential regime $E_{free}/E_{pot}<0.3$ and $\vert H_J \vert/\phi^2 <0.01$ . This enhancement is consistent with the expectation that strongly nonpotential magnetic fields are more prone to reconnection.
Recently, an empirical threshold for the ratio of current-carrying helicity to total relative helicity as $0.38 \pm 0.04$ has been proposed based on numerical simulation \citep{2026Bian}, which we failed to confirm in the present sample.

\begin{figure}[h]
\centering
\includegraphics[width=0.5\textwidth]{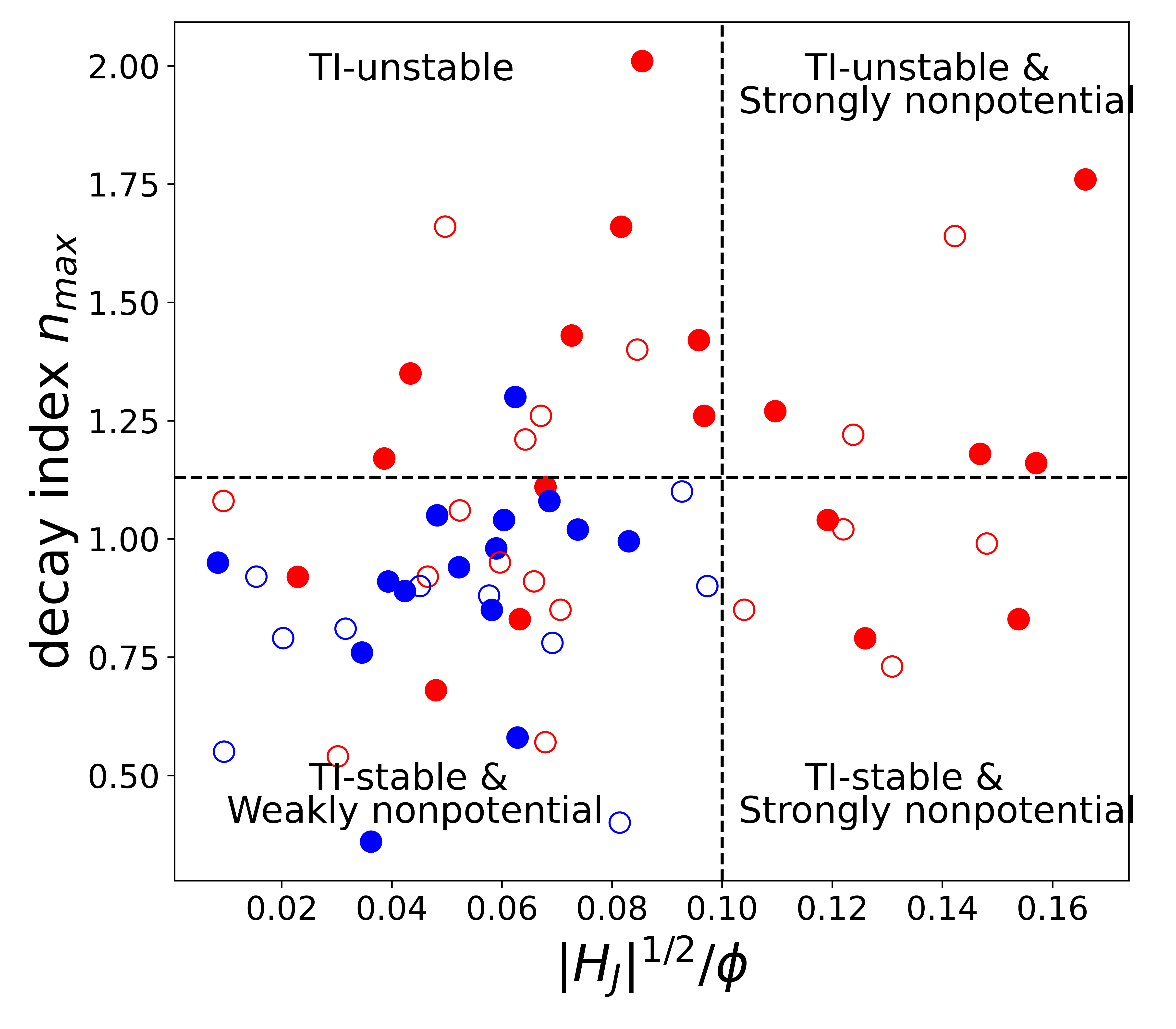}
\caption{Same as Figure \ref{KI_TI_map1} except the KI parameter replaced by the nonpotential-parameter, $ \vert H_J \vert/^{1/2}/\phi$. We adopt the same empirical thresholds used in Figure \ref{KI_TI_map1} and Figure \ref{helicity_energy}b (i.e., $n_{\rm max} = 1.13$ and $\vert H_J \vert^{1/2}/ \phi = 0.1$) to divide the space into four regimes.  
} \label{helicity_TI_map}
\end{figure}

It appears that the combination of free energy ratio $E_{free}/E_{pot}$ and the normalized current-carrying helicity above some thresholds can be good proxies for the eruptive flare event prediction. These two intensive parameters have also been analyzed to study pre-flare conditions in recent statistical studies \citep{2023Duan, 2025Thalmann}. Actually, $\vert H_J \vert/\phi^2$ could be a better parameter for this purpose as it shows a good correlation with $E_{free}/E_{pot}$, and predicts a few more eruptive events despite low $E_{free}/E_{pot}$. On the other hand, neither $\vert H_R\vert/\phi^2$ nor  $\vert H_J/H_R \vert$ shares this capability of separating the eruptive from the confined flares. 
We thus consider treating $\sqrt{\vert H_J \vert/}\phi$ as a representative parameter for eruption. Figure \ref{helicity_TI_map} presents a scatter plot in the same manner as Figure \ref{KI_TI_map1}, but with the KI parameter replaced by  this nonpotentiality parameter. In this plot, the eruptive events (red circles) are mostly found in the three quadrants corresponding to either TI-unstable regime (11 events) or TI-stable \& strongly nonpotential ARs (7 events), with 6 events in TI-unstable \& strongly nonpotential regime. In contrast, nearly all confined events  are only in the lower left quadrant representing TI-stable MFRs embedded in near-potential ARs. One exceptional event is the SOL2024-03-23T13:51 M5.4 flare that is TI-unstable but actually confined (the only blue point in the TI-unstable regime). 
This event occurs in a double-arc-shaped MFR (see  Figure \ref{special_case1}a in Appendix) with high $n$ found at the two apexes, thus classified as TI-unstable. However, a low $n$ applies to the middle portion of the arcs, where the strong strapping field suppresses its eruption. Therefore this event could have been classified as a TI-stable event if our criterion is generalized to accommodate such complexity. Except for this event, all other confined events are correctly predicted by the combined criterion of $n$ and/or $\sqrt{\vert H_J \vert/}\phi$. 

Note that these two conditions for eruptiveness, TI parameter and normalized current-carrying helicity, are not mutually exclusive. An event from high non-potential AR can erupt despite being stable against TI, and a TI-unstable MFR can erupt even when embedded in a nearly potential AR.  Together, these two conditions account for 24 eruptive events. There are still 12 eruptive events that satisfy neither conditions, which may suggest additional factors for their eruption. Our intention to plot Figure \ref{helicity_TI_map} is to examine whether the non-potential parameter can provide a better separation for eruptive events from the confined ones than the TI parameter alone.  Specifically, 67\% of the eruptive events satisfy either $n \geq 1.13$ or $\sqrt{\vert H_J \vert}/\phi \geq 0.1$, and 96\% of the confined events satisfy both $n < 1.13$ and $\sqrt{\vert H_J \vert}/\phi < 0.1$. Therefore, this combined criterion is more powerful to predict the confined flares. The helicity, $H_J$, in this figure is computed for the entire AR. If this parameter is computed locally for the isolated MFR regime, the relative helicity is generally not strictly gauge-invariant unless the boundary condition is satisfied (e.g., $\textbf{B}\times \textbf{n} = B_p \times \textbf{n}$).

\subsection{Discriminating Power among Parameters}

Table \ref{discriminating_power} ranks the parameters by their discriminating power in separating between eruptive and confined events. The discriminating power is quantified using two metrics, Cohen's d (measurement of the standardized difference between two groups) and Area Under ROC Curve (AUC, a metric quantifies the probability that a parameter as a classifier can separate two groups), with the ranking based on Cohen's d in descending order. The combination of Cohen's d and the AUC can help to describe the discriminating power of all derived parameters. Although the ranking orders are not identical for the two metrics, the top two parameters are consistent: the current helicity normalized by AR flux, $\vert H_J \vert/\phi^2$, and the TI parameter, decay index, $n_{max}$. The values for these two parameters, 0.88--0.9 in Cohen's d and 0.73--0.75 in AUC, indicate moderate-to-strong discriminating power and are substantially higher than those of the other parameters. The KI parameter, $\left|T_{w} \right|$, ranks near the lower end, with a Cohen's d of 0.129 and an AUC of 0.54, indicating weak discriminating power and predictive capability.

In addition, AR flux shows a moderate separation in mean values (Cohen's d = 0.506), but its AUC value (0.34) is the lowest among all parameters, indicating weak and effectively reversed discriminating capability. This suggests that eruptive and confined events differ in their AR size, but not in a monotonic sense. Combined with the results from Section \ref{subsec: nonpontentiality}, this implies that more extended (and less nonpotential) ARs tend to host confined events, and more compact (and more highly potential) ARs are more likely to produce eruptive events.

We also notice several interesting properties related to the ranking. (1) Different definitions of the TI and KI parameters lead to different discriminating power. For example, the maximum decay index along MFR field lines $n_{max}$, adopted in this study, ranks higher than the mean decay index at the MFR axis $n_{axis}$, as used by \cite{2018Jing}. (2) Current helicity ($\vert H_J \vert$), relative magnetic helicity ($\vert H_R \vert$) and free magnetic energy ($E_{free}$) all quantify the nonpotentiality of ARs, but reflect different physical aspects, resulting in markedly different rankings. (3) Normalized nonpotentiality parameters generally rank higher than their extensive counterparts. 

As some parameters may be correlated, we also calculate an additional metric, Least Absolute Shrinkage and Selection Operator (LASSO), to quantify the independent contributions of different parameters. Among the key physical parameters, $H_j/{\phi}^2$ and $n_{max}$ show the highest LASSO coefficients (3.00 and 1.93, respectively), indicating that they provide the strongest independent contributions. This is consistent with their high rankings in both Cohen's d and AUC values, reinforcing their strong and robust discriminating power. The remaining parameters show smaller or even negative coefficients, in decreasing order: $T_w$ (0.83), $H_R/{\phi}^2$ ($-$0.03), $T_w \cdot n_{max}$ ($-$1.17) and $E_{free}/E_{pot}$ ($-$1.42), indicating that they provide limited or redundant information once correlation are taken into account.    

\begin{table}
\begin{center}
\caption{Ranking of Discriminating Power of Parameters}\label{summary}
\begin{tabular}{ccccccccccccc}
\tableline
\tableline
Parameter & $\vert H_J \vert/ \phi_{AR}^2$&$n_{max}$ & $E_{free} /E_{pot}$ & $n_{axis}$ & $\phi_{AR}$ & $\vert H_R \vert/ \phi_{AR}^2$ & $H_R$ & $E_{free}$ & $\left|T_{w} \right|$&$z_{ap}/x_f$ & $H_J$ & Axis $T_w$ \\
\tableline
Cohen’s d\tablenotemark{a}& \textcolor{red}{0.901}&\textcolor{red}{0.885}& \textcolor{orange}{0.658}& \textcolor{orange}{0.652}&\textcolor{orange}{0.506} & \textcolor{cyan}{0.304}& \textcolor{cyan}{0.247}& \textcolor{cyan}{0.206}& \textcolor{cyan}{0.129}& \textcolor{cyan}{0.036} & \textcolor{cyan}{0.035}&\textcolor{cyan}{0.028} \\
AUC\tablenotemark{b}      & \textcolor{orange}{0.75}&\textcolor{orange}{0.73}  & \textcolor{cyan}{0.66} & \textcolor{cyan}{0.65} &0.34 & \textcolor{cyan}{0.67}&  \textcolor{cyan}{0.57} & 0.45&  \textcolor{cyan}{0.542} & 0.47 & \textcolor{cyan}{0.59}& 0.5 \\ 
\tableline
\tableline
\end{tabular}

\tablenotetext{a}{measure the difference between two groups by assuming a roughly normal distribution. $d > 0.8$: \textcolor{red}{strong}; $ 0.8 > d > 0.5$: \textcolor{orange}{moderate}; $ d < 0.2$: \textcolor{cyan}{weak}.}
\tablenotetext{b}{Area Under the ROC Curve, $AUC > 0.8$: \textcolor{red}{strong}; $ 0.8 > AUC > 0.7$: \textcolor{orange}{moderate}; $ 0.7 > AUC > 0.5$: \textcolor{cyan}{weak}; $ AUC < 0.5$: random.}
\label{discriminating_power}
\end{center}
\end{table}

\begin{figure}[h]
\centering
\includegraphics[width=0.8\textwidth]{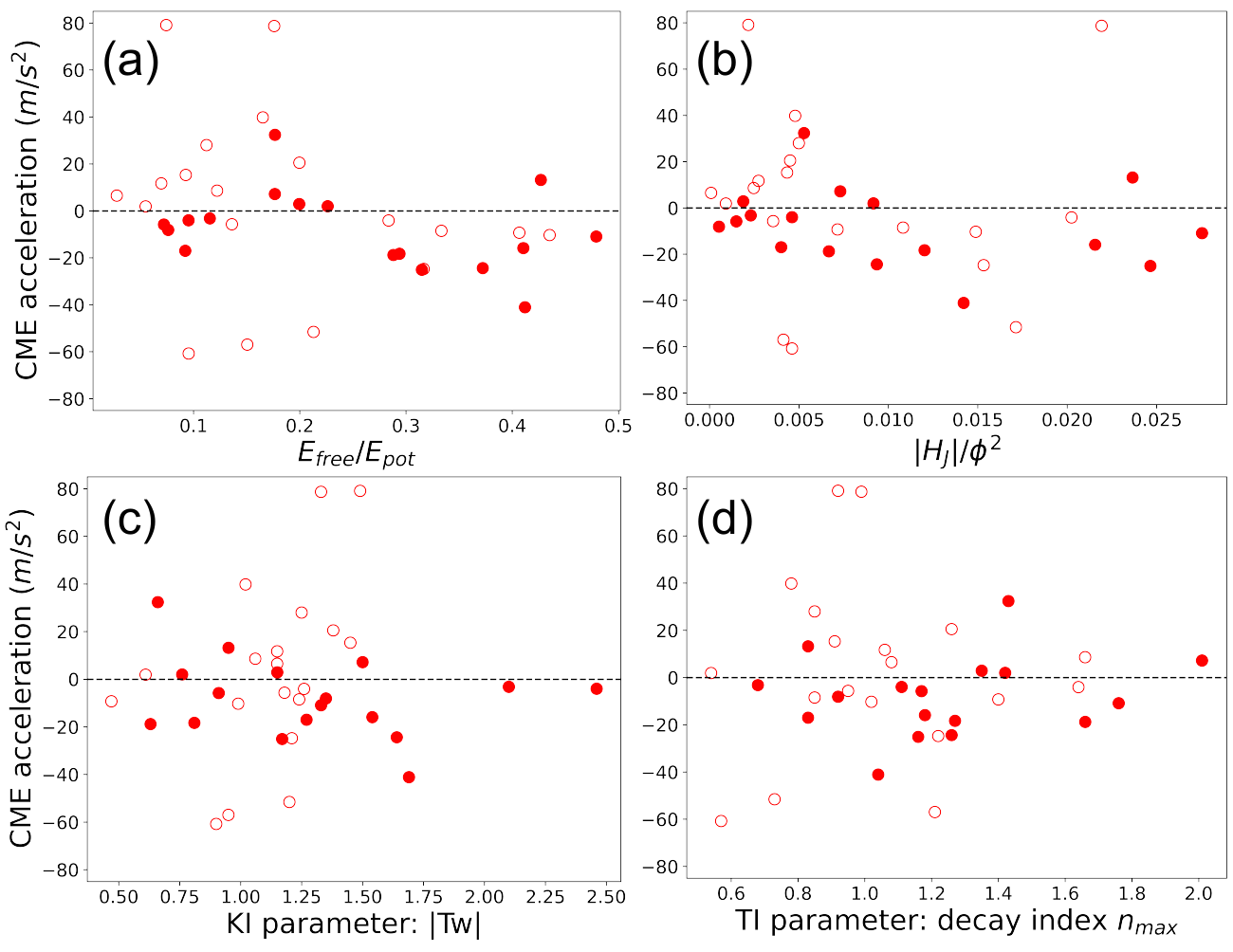}
\caption{The scatter diagrams of CME acceleration vs $E_{free}/E_{pot}$ (a), $\left| {H_J} \right|/\phi^{2}$ (b), the magnetic twist $ \left| {T_w} \right|$ (c), and the decay index $n_{max}$. Same as Figure \ref{KI_TI_map1}, the events with (without) PBs are filled (unfilled).}\label{CME_KI_TI}
\end{figure}

\subsection{CME Acceleration}

In the preceding sections, we compared the physical parameters of two groups, eruptive events and confined events, to identify the conditions that distinguish them. In this section, we focus on a subset of 37 eruptive events, all of which have identified MFRs and available estimates of CME acceleration, to examine whether their physical parameters show a clear positive correlation with CME acceleration. The values of CME accelerations for these events are derived  from 2nd-order fits to the CME height-time measurements. 

Given the data availability from 2011 to 2024, we primarily adopt the acceleration measurements from the SOHO/LASCO CME catalog\footnote{\url{https://cdaw.gsfc.nasa.gov/CME_list/}}. For events whose CME acceleration estimations are labeled as ``uncertain''  due to poor height measurement or limited measurement frames, we consult the  STEREO/SECCHI COR2 CME catalog\footnote{\url{http://spaceweather.gmu.edu/seeds/secchi.php}} \citep[see also][]{2008Olmedo_seeds}. When the CME acceleration measurements are available from both STEREO-A and STEREO-B and having comparable values, we use their average. When the STEREO-A and STEREO-B estimations differ substantially, or only one STEREO satellite measurement is available, we retain the estimation from SOHO/LASCO. 

Figure \ref{CME_KI_TI} shows the scatter plots of estimated CME acceleration versus four parameters: (a) $E_{free}/E_{pot}$, (b) $\left| {H_J} \right|/\phi^{2}$, (c) the KI parameter $\left|{T_w}\right|$ and (d) the TI parameter $n_{max}$. In general, no significant correlation is found between the CME acceleration and any of these parameters. Among these eruptive events, the mean absolute CME acceleration is 17.3 $m/s^{2}$ for the 18 events with PBs, compared to 24.1 $m/s^2$ for the 19 events without PBs. Events with PBs are more concentrated in the low to moderate acceleration range (within a range of [-30, 30] $m/s^{2}$), whereas events without PBs exhibit a broader distribution of accelerations. Specifically, as shown in Figure \ref{CME_KI_TI}a, when $E_{free}/E_{pot} <0.3$, events with PBs generally show lower absolute CME acceleration values than those events without PBs.

\section{Summary and Discussion}\label{3}

This study extends previous statistical work (e.g., \citealt{2018Jing}) through a larger sample and an expanded set of parameters to assess the roles of ideal MHD instabilities, the overall nonpotentiality of ARs and pre-flare magnetic reconnection in determining whether a flare is confined or eruptive. Compared to earlier studies, we explicitly include PFR into the analysis. This is motivated by previous findings that some eruptive events occur in the MHD TI/KI-stable regime and therefore cannot be fully explained by TI and KI alone, suggesting that PFR may provide an additional mechanism. The key results are summarized as follows.

\begin{enumerate}
\item  In contrast to the TI, the KI is much less effective at distinguishing eruptive from confined events, playing at most a minor role. 17 out of the 18 events in TI-unstable ($n>1.13$) regime are eruptive, and one is confined. In contrast, among the 19 KI-unstable ($T_w>1.3$) events, 12 events are eruptive. 

\item PFR, while capable of accounting for some eruptive events in MHD confined regime, is neither necessary nor sufficient for producing eruptive events, and plays a limited role in distinguishing eruptive from confined events. Among the 30 events stable against both KI and TI, three out of twelve eruptive events are associated with PBs, partially supporting the role of PFR in triggering eruptions, whereas the remaining nine eruptive events show no identifiable PBs. 

\item PFR occurs more frequently in TI-, KI- and KI/TI-unstable regimes, but is strongly associated with eruptive outcomes only when the TI criterion is satisfied. It suggests that TI provides the primary condition for eruptions, while PFR plays a secondary, facilitating role that acts in conjunction with TI.

\item  Strong magnetic nonpotentiality (in the form of current-carrying helicity normalized by AR flux, $\vert H_J \vert/\phi^2$) and weak overlying confinement  (represented by TI and quantified by the maximum decay index of the MFRs) show the strongest discriminating power in distinguishing eruptive from confined events. 

\item The combination of two parameters, $n$ and $\sqrt{\vert H_J \vert}/\phi$, provides a powerful condition for the eruptiveness. In our sample, 67\% (24 of 36) of the eruptive flares have either $n \geq 1.13$ or $\sqrt{\vert H_J \vert}/\phi \geq 0.1$, and 96\% (24 of 25) of the confined flares satisfy the condition of $n < 1.13$ and $\sqrt{\vert H_J \vert} < 0.1$. Therefore, this criterion is more powerful to predict the confined flares.

\item Once correlation among parameters are accounted for, the remaining parameters contribute limited or redundant information. However, subtle yet systematic differences between eruptive and confined events can still be identified from them. For example, eruptive events tend to have higher height-to-distance ratio of MFRs than confined events. Eruptive (confined) events tend to take place in more compact (extended) and more (less) nonpotential ARs. 

\item None of the considered parameters derived for the pre-flare conditions shows a statistically significant positive correlation with CME speed.

\end{enumerate}

Strong nonpotentiality of ARs and weak constraint by the overlying strapping field are critical for the initiation and successful development of eruptions, and thus are robust indicators for flare eruptiveness. KI, although capable of initiating the expansion and deformation of MFRs, is generally insufficient to drive them into the corona and heliosphere to become CMEs; accordingly, its discriminating power is limited. Both results are consistent with our expectations and align with earlier findings. 
However, the limited role of PFR in predicting eruptive from confined events is somewhat unexpected. In our sample, PFR accounts for only three of the twelve eruptive events in the MHD confined regime, leaving the remaining nine events unexplained. 

This discrepancy may be partly attributed to limitations in our observational diagnostics. Specifically, the identification of PFR relies on the detection of PBs, and the spatial association criteria adopted here are better suited for  identifying localized PBs around the flaring PIL. This makes the approach more sensitive to low-lying tether-cutting reconnection occurring near the PIL. Consistent with this, the three eruptive events with identified PBs in the confined regime are all associated with tether-cutting reconnection (see Appendix B). By contrast, breakout reconnection typically occurs at higher altitudes and tends to produce PBs that are more spatially extended and less predictable in location and morphology, making it difficult to define a consistent identification criterion for statistical analysis. Consequently, the significance of PFR may have been underrepresented or not fully captured in our analysis. 

The limited role of PFR in this study may also reflect its main contribution in initiating eruptions rather than determining their ultimate eruptive outcome. PFR, such as tether-cutting reconnection, can trigger flare onset and drive the initial upward motion of the MFR into the low corona. However, whether an event ultimately remains confined or evolves into a CME is also largely governed by the subsequent evolution and acceleration of the MFR in the higher corona. In this sense, flare eruptiveness depends not only on the pre-flare initiation process but also on the sustained driving mechanisms at later stages. As a result, the use of PFR to determine whether a flare is eruptive or confined may be only partially justified. 

This perspective also reflects a broader limitation of our analysis, namely,  the simplified treatment of the temporal and spatial complexities in solar eruptions. From a temporal perspective, the parameters derived from a single pre-flare magnetic-field snapshot may not fully capture the dynamic evolution of MFRs, which can substantially influence the eruption outcome. Spatially, we reduce the complexity of MFRs and their surrounding magnetic fields to the TI and KI conditions represented by the maximum decay index and the average twist number of the MFR, respectively, thereby neglecting the effects of spatially nonuniform strapping fields and irregular MFR morphology.

Despite these limitations, this statistical study represents a step forward toward a better understanding of the connection between complex pre-flare magnetic environment and the flare eruptiveness. The role of magnetic reconnection in driving solar eruptions warrants further investigation. Unlike the relative clean trend in laboratory experiments,  solar corona environment is much more complicated, therefore,  TI/KI combined do not uniquely decide the properties of eruptions.   Future advances may include establishing more robust characterization for breakout reconnection, incorporating time-dependent magnetic field evolution, and including quantitative measures of reconnection during the flare (e.g., the reconnection rate). In addition, detailed case studies of special events, such as confined events in the MHD TI/KI-unstable regime, as well as eruptive events found in the TI/KI-stable regime or TI-stable/near-potential regime, combined with data-driven MHD simulations, would help to place the statistical results into a clearer physical context and provide further insights into the mechanisms governing the flare eruptiveness.

\section*{Data Availability}
The event-by-event diagnostic results and the magnetic flux rope (MFR) parameter catalog for all 68 major solar flares analyzed in this study are publicly available on Zenodo: \url{https://doi.org/10.5281/zenodo.23145233}. 

\begin{acknowledgments}
We are grateful to the referee who provided value comments leading to substantial improvement of this paper. 
We thank the NASA SDO team for HMI and AIA data. HMI and AIA are instruments on board SDO, a mission for NASA's Living with a Star program. We thank NOAA for providing the GOES 15 and GOES-R series data. This study is supported by NASA grants 80NSSC21K1671, 80NSSC21K0003, 80NSSC23K0406, 80NSSC24M0174, and NSF grants AGS-2206424, 2300341 and 2309939. 

\end{acknowledgments}

%
\facilities{SDO, SOHO, STEREO, GOES 15 and 16}

\software{Sunpy, IDL, Paraview}


\appendix

\section{Flare events and associated parameters}
Here is the full list of the flare events, their time and location as well as the associated parameters used in this study. All parameters are derived from the potential and NLFFF extrapolation from a vector magnetogram chosen before the flare. 

\startlongtable
\begin{deluxetable*}{ccccccccccccccc} 
\scriptsize
\setlength{\tabcolsep}{2pt}
\tablecaption{List of Events and Associated Preflare Properties 
 \label{flare_event_list}}
\tablewidth{0pt}

\tablehead{
\colhead{No.} &\colhead{Flare peak time} & \colhead{Flare} & \colhead{NOAA} & \colhead{Location} &\colhead{E/C\tablenotemark{a}} & \colhead{PB\tablenotemark{b}} &\colhead{$n_{max}$\tablenotemark{c}} &\colhead{$\left| n\right|$ \tablenotemark{d}} &\colhead{$T_{w}$\tablenotemark{e}} &\colhead{$\left| T_w\right|$\tablenotemark{f}} & \colhead{$\phi$\tablenotemark{g}} & \colhead{$E_{f}/E_{p}$\tablenotemark{h}}
&\colhead{$H_{v}$\tablenotemark{i} } & \colhead{$H_{j}$\tablenotemark{j} }\\
\nocolhead{No.} &\nocolhead{Flare peak time} & \colhead{Class} & \colhead{AR} & \colhead{} &\colhead{} & \colhead{} &\colhead{} &\colhead{} & \colhead{}& \colhead{} & \colhead{ ($10^{22}$Mx)} & \colhead{} &\colhead{($10^{40}$Mx$^2$)} & \colhead{($10^{40}$Mx$^2$)}
}
\startdata
1&2011-02-13T17:38 & M6.6 & 11158 & S20E04 & E & Y & 1.42 & $0.83\pm 0.11$ & 0.84 & $0.76\pm 0.05$ & 1.94 &  0.23 & 2.19 & 0.86 \\
2&2011-02-15T01:56 & X2.2 & 11158 & S20W10 & E & Y & 1.27 & $0.61\pm 0.32$ & 0.99 & $0.81\pm 0.21$ & 2.73 &  0.29 & 8.95 & 2.24 \\
3&2011-03-09T23:23 & X1.5 & 11166 & N08W09 & C & Y & 1.08 & $0.40\pm 0.29$ & 1.47 & $1.22\pm 0.17$ & 3.62 &   0.16 & 3.76 & 1.54 \\
4&2011-07-30T02:09 & M9.3 & 11261 & S20W10 & C & N & 0.88 & $0.35\pm 0.30$ & -1.56& $1.26\pm 0.15$ & 2.56 &  0.18 & -2.31 & -0.54 \\
5&2011-08-03T13:48 & M6.0 & 11261 & N16W30 & E & N & 1.02 & $0.59\pm 0.21$ & 1.50 & $0.99\pm 0.17$ & 2.65 &  0.44 & 12.08 & 2.61 \\
6&2011-08-04T03:57 & M9.3 & 11261 & N17W37 & E & Y & 1.04 & $0.57\pm 0.20$ & 1.59 & $1.69\pm 0.19$ &3.24 &  0.41 & 17.11 & 3.74 \\
7&2011-09-06T01:50 & M5.3 & 11283 & N14W07 & E & N & 1.66 & $0.62\pm 0.26$ & 1.63 & $1.06\pm 0.27$ & 2.34 & 0.12 & 3.14 & 0.34  \\
8&2011-09-06T22:20 & X2.1 & 11283 & N14W18 & E & Y & 1.43 & $0.74\pm 0.27$ & 0.80 & $0.66\pm 0.04$ & 2.25 & 0.18 & 4.06 & 0.67  \\ 
9&2011-09-07T22:38 & X1.8 & 11283 & N14W28 & E & Y & 2.01 & $0.81\pm 0.38$ & 2.16 & $1.50\pm 0.33$ & 2.52 & 0.18 & 1.16 & 1.08  \\
10&2011-10-02T00:50 & M3.9 & 11305 & N10W14 & C & N & 0.40 & $0.21\pm 0.10$ &-1.33 & $1.01\pm 0.11$ & 1.28&  0.12 & -7.31& -0.27 \\
11&2012-01-23T03:59 & M8.7 & 11402 & N28W21 & E & N & 0.85 & $0.48\pm 0.17$ &-1.71 & $1.25\pm 0.15$ & 7.61&  0.11 & 19.60 & 7.24 \\
12&2012-03-07T00:24 & X5.4 & 11429 & N17E31 & E & Y & 0.79 & $0.47\pm 0.15$ &-1.39 & $1.12\pm 0.11$ & 5.15&  0.43 & -89.98 & -10.52 \\
13&2012-03-09T03:53 & M6.3 & 11429 & N15W03 & E & Y & 0.83 & $0.34\pm 0.16$ & 1.57 & $0.95\pm 0.24$ &4.38&  0.43 & -52.46 & -11.37 \\
14&2012-03-10T17:44 & M8.4 & 11429 & N16W24 & E & Y & 1.76 & $0.72\pm 0.40$ &-1.60 & $1.33\pm 0.16$ & 4.41&   0.48 & -54.97& -13.42 \\
15&2012-05-10T04:18 & M5.7 & 11476 & N10E08 & C & N & 0.92 & $0.74\pm 0.25$ &-1.42 & $1.24\pm 0.11$ & 5.77&  0.06 & -0.29 & -0.20 \\
16&2012-07-02T10:52 & M5.6 & 11515 & S17E08 & E & Y & 0.68 & $0.26\pm 0.17$ &-2.39 & $2.10\pm 0.18$ & 3.78&  0.12 & -24.22 & 0.82 \\
17&2012-07-05T11:44 & M6.1 & 11515 & S17W37 & C & Y & 0.58 & $0.07\pm 0.26$ & 0.87 & $0.54\pm 0.09$ &5.66&  0.17 &-40.73& -3.16 \\
18&2013-04-11T06:55 & M6.5 & 11719 & N09E12 & E & Y & 0.92 & $0.36\pm 0.26$ &-1.98 & $1.35\pm 0.27$ & 2.15&  0.08 & -1.49 & 0.06 \\
19&2013-10-24T00:30 & M9.3 & 11877 & S09E10 & E & Y & 0.83 & $0.20\pm 0.25$ & 2.19 & $1.27\pm0.42$ & 3.77&  0.09 & 28.55 & 1.42 \\
20&2013-11-01T19:53 & M6.3 & 11884 & S12E01 & C & Y & 0.76 & $0.51\pm 0.13$ & 1.45 & $1.18\pm 0.08$ & 3.68&  0.11 & 11.07 & 0.41 \\
21&2013-11-03T05:22 & M5.0 & 11884 & S12W17 & C & N & 0.79 & $0.35\pm 0.17$ & 1.95 & $1.07\pm 0.17$ & 3.33&  0.08 &  5.77 &-0.11 \\
22&2013-11-05T22:12 & X3.3 & 11890 & S12E44 & E & Y & 1.17 & $0.18\pm 0.41$ & 1.15 & $0.91\pm 0.07$ & 7.22&  0.07 & 13.97 & 1.95 \\
23&2013-11-08T04:26 & X1.1 & 11890 & S12E13 & E & N & 1.06 & $0.34\pm 0.38$ & 1.14 & $1.15\pm 0.12$ & 4.74&  0.07 & 18.19 & 1.54 \\
24&2013-11-10T05:14 & X1.1 & 11890 & S14W13 & E & **& 1.08 & $0.53\pm 0.18$ & 1.43 & $1.15\pm 0.10$ & 4.72&  0.03 & -7.06 & 0.05 \\
25&2013-12-31T21:58 & M6.4 & 11936 & S16W35 & E & Y & 1.35 & $0.91\pm 0.24$ &-1.39 & $1.15\pm 0.11$ & 4.38&  0.20 & -2.85 &-0.90 \\
26&2014-01-07T10:13 & M7.2 & 11944 & S15W11 & E & N & 0.91 & $0.39\pm 0.26$ &-1.97 & $1.45\pm 0.25$ & 8.41& 0.09& 186.11&7.66\\
27&2014-01-07T18:32 & X1.2 & 11944 & S12W08 & E & N & 0.57 & $0.28\pm 0.19$ & 1.20 & $0.90\pm 0.17$ & 8.23& 0.10 & 187.00& 7.82 \\
28&2014-02-02T09:31 & M4.4 & 11967 & S10E14 & C & Y & NA  &  NA & NA  &  NA & 7.79&  0.18 &-31.65 &-5.93 \\
29&2014-02-04T04:00 & M5.2 & 11976 & S14W06 & C & Y & 0.85 & $0.41\pm 0.19$ &-2.28 & $1.17\pm 0.37$ & 7.93 & 0.15 &-42.92 &-5.32 \\
30&2014-03-29T17:48 & X1.0 & 12017 & N11E05 & E & N & 1.64 & $0.92\pm 0.29$ & 1.64 & $1.26\pm 0.17$ & 1.83 & 0.28 & 7.77  & 1.69 \\
31&2014-09-10T17:45 & X1.6 & 12158 & N11E05 & E & ** &0.73 & $0.40\pm 0.19$ &-1.49 & $1.20\pm 0.14$ & 2.85 & 0.21 &-35.79 &-3.47 \\
32&2014-09-28T02:58 & M5.1 & 12173 & S13W23 & E & ** & NA  & NA & NA  & NA  & 6.65 & 0.19 &-32.62 &-7.89 \\
33&2014-10-22T01:59 & M8.7 & 12192 & S13E21 & C & N & 0.78 & $0.47\pm 0.17$ &-1.36 & $1.02\pm 0.16$ & 14.30& 0.17 &-499.45&-24.47\\
34&2014-10-24T07:37 & M4.0 & 12192 & S19W05 & E & Y & NA   & NA & NA  & NA & 14.98& 0.19 &-655.47&-27.72\\
35&2014-10-24T21:41 & X3.1 & 12192 & S22W21 & C & Y & 1.04 & $0.45\pm 0.22$ &-2.94 & $1.83\pm 0.52$ & 15.79& 0.17 &-710.08&-22.75\\
36&2014-10-25T17:08 & X1.0 & 12192 & S10W22 & C & Y & 0.98 & $0.46\pm 0.23$ &-2.99 & $2.30\pm0.39$  & 17.04& 0.17 &-751.48&-25.24\\
37&2014-11-07T17:26 & X1.6 & 12205 & N17E40 & E & Y & 1.26 & $0.69\pm 0.27$ & 2.19 & $1.64\pm 0.34$ & 4.40 & 0.37 &  4.64 & 4.54 \\
38&2014-12-04T18:25 & M6.1 & 12222 & S20W31	& C	& Y & 1.05 & $0.67\pm 0.32$ & 1.33 & $1.17\pm 0.09$ & 4.20 & 0.04 &-20.25 & -1.03\\ 
39&2014-12-17T04:51 & M8.7 & 12242 & S18E08 & E	& N	& NA   & NA & NA & NA& 5.23 & 0.25 &  2.97 & 5.16 \\
40&2014-12-18T21:58 & M6.9 & 12241 & S11E15	& E	& N & 1.21 &$0.49\pm 0.38$ & 1.13 & $0.95\pm 0.09$  & 3.50 & 0.15 &24.07&1.27 \\ 
41&2014-12-20T00:28 & X1.8 & 12242 & S19W29	& E	& N &0.85 & $0.48\pm 0.22$ & 1.46 & $1.24\pm 0.13$ &	7.42 & 0.33 & 103.67& 14.89\\ 
42&2015-03-11T16:22 & X2.1 & 12297 & S17E22 & E & Y & 1.18 &$0.66\pm 0.30$ & 2.39 & $1.54\pm 0.26$ & 3.03 & 0.41 & 27.88 & 4.96 \\
43&2015-03-12T14:08 & M4.2 & 12297 & S15E06 & C & N & 0.90 &$0.42\pm 0.20$ & 1.33 & $1.03\pm 0.10$ & 2.99 & 0.29 & 16.20 & 2.12 \\ 
44&2015-06-22T18:23 & M6.5 & 12371 & N13W06 & E & Y & 1.16 &$0.71\pm 0.25$ &-1.36 & $1.17\pm 0.09$ &  4.26 & 0.31 &-78.47 &-11.18\\					
45&2015-06-25T08:16 & M7.9 & 12371 & N12W40 & E & **& 1.22 &$0.74\pm 0.34$ &-1.53 & $1.21\pm 0.12$ &    5.76 & 0.32 &-84.54 &-12.70\\				
46&2015-08-24T07:33 & M5.6 & 12403 & S14E00 & C & Y & 0.36 &$0.15\pm 0.07$ & 1.33 & $1.19\pm 0.07$ &   5.89 & 0.08 & 18.97 & 1.14 \\							
47&2015-09-28T14:58 & M7.6 & 12422 & S20W28 & C & Y & 0.95 &$0.38\pm 0.32$& -2.95 & $1.48\pm 0.55$ & 4.61 & 0.07&-19.25 &0.04 \\						
48&2017-09-04T20:33 & M8.4 & 12673 & S10W11 & E & Y &  NA   & NA &  NA   & NA &    3.48 & 0.48 &-41.71 &-6.78 \\						
49&2017-09-06T12:02 & X14.6 & 12673 & S09W34 & E & N & 0.92 &$0.56\pm 0.18$ &-1.74 & $1.49\pm 0.14$ & 5.16 & 0.07 &-34.07 &-1.44 \\ 						
50&2022-03-30T17:37 & X1.3 & 12975 & N13W31 & E & N & 1.40 &$0.65\pm 0.41$ & 0.76 & $0.47\pm 0.12$ & 3.66 & 0.41 & 0.96  & 2.40 \\
51&2022-04-21T01:59 & M9.6 & 12993 & N22E23 & E & N & NA   & NA & NA  & NA  & 10.09& 0.15 & 34.70 & 2.36 \\
52&2022-05-04T08:59 & M5.7 & 13004 & S15W07 & C & Y & 0.94 &$0.44\pm 0.23$ & 1.68 & $1.41\pm 0.18$ & 1.71 & 0.19 & -0.55 & 0.20 \\
53&2022-05-10T13:55 & X1.5 & 13006 & S29W04 & E & Y & 1.11 &$0.54\pm 0.19$ &-3.39 & $2.46\pm 0.57$ & 3.94 & 0.10 & -14.44& -1.79\\ 
54&2022-05-19T07:19 & M5.6 & 13017 & N18E26 & C & Y & 0.91 &$0.40\pm 0.28$ & 1.43 & $1.06\pm 0.16$ & 9.45& 0.14 & 5.52  & -3.46\\ 
55&2022-10-01T20:10 & M5.8 & 13110 & N17W35 & E & N & 0.99 &$0.45\pm 0.28$ & 2.96 & $1.33\pm 0.39$ & 0.98 & 0.18 & 2.84  &  0.53\\ 
56&2022-10-02T02:21 & M8.7 & 13110 & N17W39 & E & N & 1.26 &$0.60\pm 0.19$ & 1.94 & $1.38\pm 0.22$ & 2.31 & 0.20 & 2.49  &  6.05\\ 
57&2023-02-07T23:05 & M6.4 & 13213 & N30E03 & C &** & 1.10 &$0.52\pm 0.32$ & 1.80 & $1.36\pm 0.22$  & 3.01 & 0.19  & -9.31 & -1.95\\ 
58&2023-02-11T15:48 & X1.1 & 13217 & S10E39 & C &** & 0.90 &$0.28\pm 0.25$ &-0.72 & $0.49\pm 0.09$ & 9.87& 0.13 & 2.98  &  4.97\\
59&2023-02-28T17:50 & M8.6 & 13234 & N27W29 & E &** & NA   & NA & NA  & NA & 5.56 & 0.11 & -16.53  & -2.76\\ 
60&2023-05-03T10:45 & M7.3 & 13293 & N12E42 & C & Y & 0.89 &$0.63\pm 0.21$ &-2.17 & $1.57\pm 0.27$ & 4.64 & 0.12 & 17.89 & 0.97\\ 
61&2023-05-09T03:54 & M6.5 & 13296 & N13W26 & E & Y & 1.66 &$1.02\pm 0.37$ & 0.66 & $0.63\pm 0.13$ & 3.88 & 0.29 & 5.21 &2.51\\
62&2023-05-09T20:52 & M5.0 & 13296 & N13W36 & C & Y & 1.02 & $0.48\pm 0.23$ & 1.49 &$1.24\pm 0.11$ & 4.85& 0.29 &9.95 & 3.21\\ 
63&2023-09-21T12:54 & M8.7 & 13435 & N07E22 & E & N & 0.95 &$0.43\pm 0.39$ & 1.46 & $1.18\pm 0.12$ & 3.05 & 0.14 & 18.81 & 0.83 \\
64&2024-02-12T03:48 & M6.5 & 13576 & S13W25 & C & N & 0.81 &$0.12\pm 0.39$ & 0.95 & $0.78\pm 0.10$ & 7.22& 0.10 & 11.22 & 1.31\\ 
65&2024-02-22T06:32 & X1.6 & 13590 & N17E35 & C & N & 0.55 &$0.28\pm 0.15$ &-0.96 & $0.82\pm 0.04$ & 9.10 & 0.06 & 2.61 & -0.19 \\
66&2024-02-22T22:34 & X6.3 & 13590 & N17E27 & E & N & 0.54 &$0.26\pm 0.15$ &-0.74 & $0.61\pm 0.07$ & 8.28 & 0.05 & -23.06 &-1.56\\ 
67&2024-03-10T12:13 & M7.4 & 13599 & S13W38 & C & Y & 1.00 &$0.58\pm 0.22$ & 0.55 & $0.36\pm 0.11$ & 4.24 & 0.16 & 30.36 & 3.09\\ 
68&2024-03-23T13:51 & M5.3 & 13615 & S14E06	& C	& Y & 1.30 &$0.57\pm 0.29$ &-1.98 & $1.50\pm 0.20$ & 5.78& 0.26 & 17.84  & 3.26\\						
\hline
\hline
\enddata

\tablecomments{Flare events without consistent MFR identifications are marked with ``NA'' in columns $n_{max}$, $\left| n\right|$, $T_w$ and $\left| T_w \right|$. }
\tablenotetext{a}{Flare type: E--eruptive, C--confined. The CME association is estimated from temporal evolutions based on SDO/AIA observations, as well as SOHO/LASCO and and STEREO CME catalogs.}
\tablenotetext{b}{Precursor Brightening (PB) before the flare onset, events with uncertain PB features are marked with ``**''.}
\tablenotetext{c}{Maximum value of decay index $n$ along MFR-associated field lines. This value is used in Figure \ref{KI_TI_map1}.}
\tablenotetext{d}{Average decay index $n$ with $\pm 1 \sigma$ uncertainty for the identified MFR field lines.}
\tablenotetext{e}{Mean unsigned magnetic twist with $\pm 1 \sigma$ uncertainty from identified MFR field lines, which is adopted in Figure \ref{KI_TI_map1} (see Section \ref{sec:TI_KI_parameters}.}
\tablenotetext{f}{Maximum signed magnetic twist from identified MFR field lines, used as a proxy of the MFR axis.}
\tablenotetext{g}{$\phi$ is the unsigned magnetic flux within the field of view of the input HMI/SHARP magnetograms.}
\tablenotetext{h}{Ratio between the free magnetic energy and the potential energy.}
\tablenotetext{i}{\textbf{$H_R$} is the relative magnetic helicity (see Section \ref{sec:additional_mag_parameters}).}
\tablenotetext{j}{\textbf{$H_J$} is non-potential current-carrying helicity (see Section \ref{sec:additional_mag_parameters}).}
\end{deluxetable*}

\section{Special Cases}

\begin{figure}[h]
\centering
\includegraphics[width=0.6\textwidth]{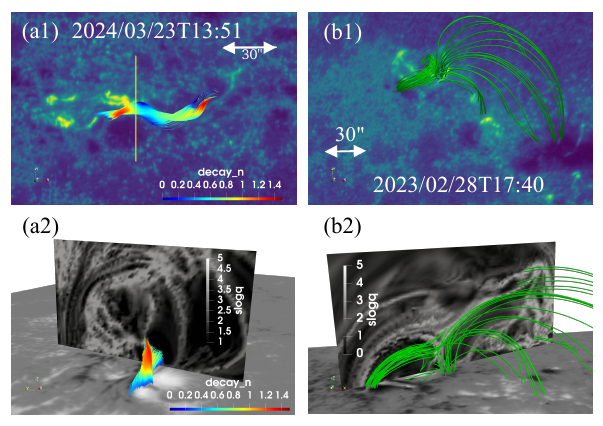}
\caption{Magnetic structure of exceptional cases.The top panels show field lines selected for representing MFRs near flaring sites over the AIA 1600 \r{A} images. The bottom panels show perspective views of the $Q$ factor on a vertical slice intersecting the magnetic field lines. (a) The 2024/03/23T13:51 flare involves  an FR with high $n_{\rm max}$ and high $T_w$. See the field lines colored according to the decay index $n$. In the middle section of the FR, $n$ is so low that it could be confined.
(b) The 2023/02/28T12:23 flare occurred in a spine-facn structure. Since no coherent MFR is found near the flaring site, we excluded it from the present study. However, a null-point reconnection could have allowed it to erupt.} \label{special_case1}
\end{figure}

As this work is primarily statistical, we examine only a few special events in detail to illustrate the complexity and diversity of magnetic field configurations across the sample. Figure \ref{special_case1}a shows the MFR structure of the SOL2024-03-23T13:51 flare, the only confined event in the KI/TI-unstable regime in Figure \ref{KI_TI_map1}, and also the only confined event in the TI-unstable regime in Figure \ref{helicity_TI_map}. The MFR has a double-arc shape, and the decay index varies significantly along the MFR reaching high values ($n_{max} =1.3$) at the two apexes, with a mean twist number of $\left|T_{w} \right| = 1.5$. Therefore it amounts to both TI/KI unstable condition in our criteria. However, in the middle portion of the MFR, the field lines are concave inward and correspond to a much lower decay index ($n = 0.87$). This suggests that  the eruption may be suppressed due to the irregular geometry of the MFR and the influence of non-uniform strapping field. If this case is regarded as effectively TI-stable, then the success rate for predicting confined events increases to 100\% in Figure \ref{helicity_TI_map}.

Figure \ref{special_case1}b shows a special case of a circular ribbon flare in a spine-fan magnetic topology. In such configurations, breakout reconnection at a magnetic null point located above the spine-fan structure can trigger magnetic eruptions  \citep{2008Lynch}. In this case, no clear MFR is identified within or around the spine-fan structure, although a group of low-lying sheared magnetic field lines ($\left|T_{w} \right| = 0.69$) is found below the null point. Continuous reconnection at the null point may transfer magnetic flux across the fan-shaped QSL, thereby facilitating eruption, consistent with the ``breakout''-type reconnection scenario. 

\begin{figure}[h]
\centering
\includegraphics[width=0.8\textwidth]{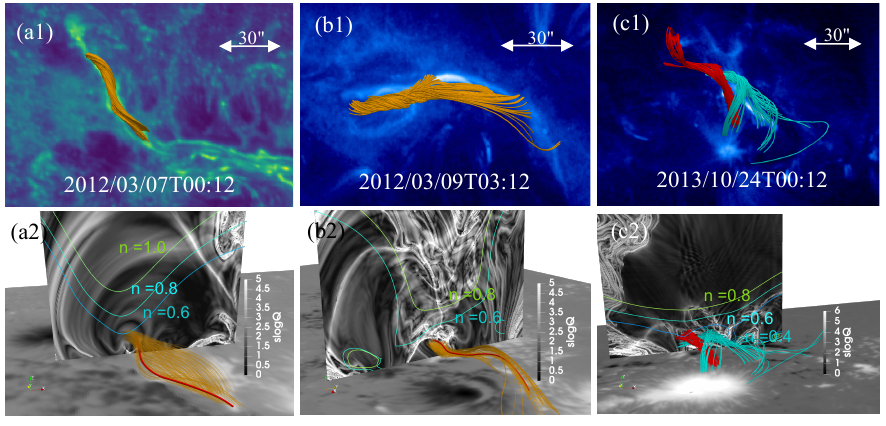}
\caption{Magnetic structure for three selected eruptive events with PBs. The top panels show the field lines selected for representing FRs near flaring sites over the AIA images. The bottom panels show perspective views of the $Q$ factor on a vertical slice intersecting the magnetic field lines. The contours of decay index n is overplotted on the vertical slices.} 
\label{special_cases}
\end{figure}

Figure \ref{special_cases} shows the three eruptive events with PBs in the TI/KI-stable regime shown in Figure \ref{KI_TI_map1}. The MFR structures are represented by the field lines , along with vertical slices of the decay index and the QSLs across the MFRs. The squashing factor $Q$ map indicates the QSLs, where magnetic connectivity varies drastically, can be preferred locations for magnetic reconnection. Enhanced QSLs are found below or close to MFRs, which lie close to the solar surface. The spatial coincidence of PBs with low-lying QSLs near the PIL, together with the presence of a pre-existing MFR, suggests that these three events are most likely triggered by tether-cutting reconnection. 
Overall, these selected cases suggest that the reconnection type associated with PBs depends on the specific interaction between the core magnetic structure and its surrounding magnetic environment, and thus can vary from case to case. Ultimately, time-dependent numerical simulations for detailed case studies are useful to further investigate the dynamics and the combined roles of magnetic reconnection and ideal MHD instabilities in solar eruptions. This statistical study utilized only a snapshot of the pre-flare magnetic topology, and more detailed time-dependent analysis will be performed in future case studies.

\bibliography{sample701}{}
\bibliographystyle{aasjournalv7}



\end{document}